\documentclass[aps,prd,nonfootinbib,twocolumn]{revtex4-2}
\usepackage{amsfonts}
\usepackage{amssymb}
\usepackage{amsmath}
\usepackage{color}
\usepackage[usenames,dvipsnames]{xcolor}
\usepackage{float}
\usepackage{accents}
\usepackage{graphicx}
\usepackage{epstopdf}
\usepackage{ulem}
\usepackage{mathtools}
\usepackage[colorlinks=true,pdfstartview=FitV,linkcolor=blue,citecolor=blue,urlcolor=blue,breaklinks=true]{hyperref}

\begin{document}

\title{{Self-dual $CP(2)$ vortex-like solitons} in the presence of magnetic
impurities}
\author{V. Almeida}
\email{vinicius.marcos@discente.ufma.br}
\author{R. Casana}
\email{rodolfo.casana@ufma.br}
\email{rodolfo.casana@gmail.com}
\affiliation{Departamento de F\'{\i}sica, Universidade Federal do
Maranh\~{a}o, {65080-805}, S\~{a}o Lu\'{\i}s, Maranh\~{a}o, Brazil.}
\author{E. da Hora}
\email{carlos.hora@ufma.br}
\email{edahora.ufma@gmail.com}
\affiliation{Coordena\c{c}\~{a}o do Curso de Bacharelado Interdisciplinar em Ci\^{e}ncia e Tecnologia,\\
Universidade Federal do Maranh\~{a}o, {65080-805}, S\~{a}o Lu\'{\i}s, Maranh\~{a}o, Brazil.}
\author{S. Krusch}
\email{S.Krusch@kent.ac.uk}
\affiliation{School of Mathematics, Statistics and Actuarial Science, University
of Kent,\ Canterbury, CT2 7FS, United Kingdom.}

\begin{abstract}
{We investigate the existence of vortex configurations in two gauged-$CP(2)$
models extended via the inclusion of magnetic impurities. In particular, we
consider both the Maxwell-$CP(2)$ and the Chern-Simons-$CP(2)$ enlarged
scenarios, separately. We choose a $CP(2)$-field configuration with a null
topological charge not only in the simplest (free) case, but also when
coupled to an Abelian gauge field. The implementation of the
Bogomol'nyi-Prasad-Sommerfield (BPS) formalism shows that the effective
models for such a configuration possess a self-dual structure which looks
like those inherent to the gauged sigma models. Therefore, when the $CP(2)$
field is coupled to the Maxwell term, the corresponding total energy
possesses both a well-defined Bogomol'nyi bound and a quantized magnetic
flux. Further, when the $CP(2)$ scenario is gauged with the Chern-Simons
action, the total electric charge is verified to be proportional to the
quantized magnetic flux. In addition, the analysis verifies that the
magnetic impurity contributes to the BPS potentials and appears in both the
models' BPS equations. Next, we introduce a Gaussian type impurity and solve
the self-dual equations via a finite-difference scheme. The resulting
solutions present a nonmonotonic behavior that flips both the magnetic and
electric fields. Finally, we discuss the topologically trivial solutions in
the limit for which the impurity becomes a Dirac $\delta $-function.}
\end{abstract}

\pacs{11.10.Kk, 11.10.Lm, 11.27.+d}
\maketitle

\section{Introduction}

\label{Intro}

{Configurations with nontrivial topology are usually achieved as the
solutions of the Euler-Lagrange equations which appear in the context of
nonlinear field theories \cite{n5}.} In this case, the nonlinearity {%
originates from} a potential term {that} promotes the spontaneous breaking
of {the original model's symmetry}. However, {the resulting second-order
Euler-Lagrange equations are highly nonlinear and} usually quite hard to
solve.

{Under {exceptional} circumstances, topological solutions can also be
obtained via a particular set of first-order differential equations, the
so-called Bogomol'nyi-Prasad-Sommerfield (BPS) ones. It is interesting to
note that these equations emerge as a result of the implementation of the
Bogomol'nyi technique, which stands for the minimization of the total energy
inherent to the field model \cite{n4}. Among the algorithms which lead to
the BPS equations, we can include the study of the conservation of the
corresponding energy-momentum tensor \cite{ano} and the On-Shell method \cite%
{onshell}. In such a scenario, the simplest BPS gauged vortices occur in the
Maxwell-Higgs \cite{n1}, Chern-Simons-Higgs \cite{cshv}, and
Maxwell-Chern-Simons-Higgs \cite{mcshv} theories.}

{More recently, {it was shown that} first-order or BPS vortices {also exist}
in a gauged scenario that describes the interaction between the Maxwell and
the $CP(2)$ fields \cite{loginov,casana}. {Besides that, some of us have
also }studied the existence of BPS vortices in the Chern-Simons-$CP(2)$ \cite%
{cscp2}, and in the Maxwell-Chern-Simons-$CP(2)$ \cite{mcscp2} models. BPS
vortices {also arise} in the context of extended {scenarios based on a }%
gauged-$CP(2)$ model, {as the Maxwell-$CP(2)$ vortices} saturated by a
nontrivial dielectric function \cite{mcp2df} and {the Maxwell-$CP(2)$
vortices} with internal structures due to the presence of an additional
scalar field \cite{mcp2is}.}

{The interactions between vortices and impurities have been observed in
various physical systems (such as condensed matter \cite{Shapoval:2010},
Bose-Einstein condensates \cite{Tung:2006}, and neutron stars \cite%
{Anderson:1975zze}), with their dynamics explored, for instance, in the
Refs. \cite{Bulgac:2013nmn, Wlazlowski:2016yoe}. More recently, a systematic
way to introduce impurities into BPS systems (which leads to the
preservation of half of the BPS equations) was developed in \cite%
{Adam:2018pvd, Adam:2019yst}. Such a procedure allows the detailed
investigation of, for example, the scattering of a kink by a kink which is
trapped by an impurity, see also the Ref. \cite{Goatham:2010dg} for an
earlier study. Moreover, an open question is how the impurities affect the
Manton-Schr\"{o}dinger-Chern-Simons model \cite{Manton:1997tg} and its
interesting vortex dynamics \cite{Romao:2004df, Krusch:2005wr}.}

On the other hand, a promising physical issue is the search for regular
solitons inherent to enlarged field theories which mimic condensed matter
phenomena. In this sense, the first studies about the existence of BPS
vortices in a Maxwell-Higgs scenario enlarged by impurities (both magnetic
and electric) were done by Tong and Wong \cite{15} and showed how the
presence of impurities affects the corresponding moduli space. Moreover, in
the Ref. \cite{16}, the authors proposed existence theorems for both
vortices and anti-vortices in the presence of magnetic impurities. In the
sequence, first-order vortices inherent to a Chern-Simons-Higgs model
extended to include impurities {were obtained in the Ref. \cite{18}}.
Furthermore, some of us investigated the interaction between a moving
Maxwell-Higgs vortex and a static magnetic impurity, see \cite{19, 20}.

{We now go further and study the occurrence of BPS vortex-like solutions in
the context of gauged $CP(2)$ models in the presence of magnetic impurities.
More specifically, we consider those topological structures engendered by
the following $CP(2)$ configuration,
\begin{equation}
\phi =\left(
\begin{array}{c}
(-1)^{k}\psi  \\
\psi ^{\ast } \\
\phi _{3}%
\end{array}%
\right) ,\quad |\phi |^{2}=h,  \label{beta1}
\end{equation}%
where $k\in \mathbb{Z}$, with $\psi \in \mathbb{C}$ and $\phi _{3}\in
\mathbb{R}$. The configuration above {is related} to the $CP(2)$ solutions
studied in {the} {refs.} \cite{loginov, casana, cscp2,mcscp2} ({in particular%
}, the configuration with $\beta _{1}=\frac{\pi }{4}+\frac{\pi k}{2}$ {%
considered} there). It is easy to verify {that} the configuration (\ref%
{beta1}) has {a} null $CP(2)$ topological charge, see {the eqs.} (\ref%
{CPch02}) and (\ref{CPch04}) {below}. {In addition,} this $CP(2)$
configuration, when minimally coupled to {the} Abelian gauge field, {%
possesses} a well-defined BPS structure (see the Secs. \ref{general0} and %
\ref{sec30} below) which supports the vortex-like solutions studied in the
Refs. \cite{loginov, casana,cscp2,mcscp2}. }

{Here, it is worthwhile to point out that the $CP(N)$ field describes
topological excitations in some cold atomic systems. For example, in the $%
s=1/2$ fermion case, the $CP(1)$ model describes the spin dynamics \cite%
{Aoki}, whereas the $CP(3)$ model can be used to study the $s=3/2$ case \cite%
{CWu}. We then expect that the $CP(2)$-configuration (\ref{beta1}) would
describe excitations of a spin-1 Bose-Einstein condensate such as vortices
or monopoles \cite{Stoof,Chang,Kasamatsu}.}

{The {present} manuscript considers {those} effective {scenarios} for the
configuration (\ref{beta1}) {which arise} from both the Maxwell-$CP(2)$ and
the Chern-Simons-$CP(2)$ models now enlarged to include a localized
impurity, which is rotationally symmetric. This way {up}, we organize our {%
work} as follows: in {the }Sec. \ref{general0}, we study the BPS structure
of the effective Maxwell-$CP(2)$ model for the configuration (\ref{beta1}). {%
In the next} Sec. \ref{sec2}, we introduce an extended Maxwell-$CP(2)$
theory saturated by an additional term which stands for the impurity itself.
{We then} look for vortex-like solutions {which minimize} the total energy
via {the implementation of} the so-called BPS prescription, {from which} we
obtain {not only} a well-defined energy lower bound, {but also} the
self-dual equations whose solutions saturate that bound. We point out the
main differences between the resulting solutions and the ones obtained
without impurities by discussing how the impurity affects the formation of
the corresponding vortices. In {the} Sec. \ref{sec30}, the BPS structure of
the effective Chern-Simons-$CP(2)$ model engendered by the configuration (%
\ref{beta1}) {is analyzed}. {In addition, the} Sec. \ref{sec3} is {dedicated
to the study of} the enlarged Chern-Simons-$CP(2)$ scenario. {Here, for the
sake of convention, }}we discuss the theoretical construction in detail by
comparing it {to} the case {considered} in {the previous} Sec. \ref{sec2}.
Finally, {the} Sec. \ref{sec4} brings our conclusions and perspectives
regarding future contributions.{\ }

\section{The effective Maxwell-$CP(2)$ model\label{general0}}

The Maxwell-$CP(2)$ model is described by the following Lagrangian density
\cite{fftt}%
\begin{equation}
\mathcal{L}=-\frac{1}{4}F_{\mu \nu }F^{\mu \nu }+\left( \nabla _{\mu }\phi
\right) ^{\dag }\left( \nabla ^{\mu }\phi \right) -U\left( \phi \right)
\text{,}  \label{1m0}
\end{equation}%
where $\phi $ is the $CP(2)$ field, which {possesses} three complex
components {which satisfy} the normalization condition $\phi ^{\dag }\phi =h$%
. The topological current density {inherent to} the $CP(2)$ field is given by%
\begin{equation}
\tau _{\mu }=\frac{1}{2\pi ih}\varepsilon _{\mu \nu \rho }\left( \mathcal{D}%
^{\nu }\phi \right) ^{\dag }\left( \mathcal{D}^{\rho }\phi \right) \text{,}
\label{CPch01}
\end{equation}%
where $\mathcal{D}^{\nu }\phi =\partial _{\mu }\phi -h^{-1}\left( \phi
^{\dag }\partial _{\mu }\phi \right) \phi $, the {resulting} topological
charge being expressed as%
\begin{equation}
\mathfrak{q}=\int d^{2}\mathbf{x~}\tau _{0}\in \mathbb{Z}\setminus \left\{
0\right\} \text{.}  \label{CPch02}
\end{equation}

Moreover, in the Eq. (\ref{1m0}), $F_{\mu \nu}=\partial _{\mu }A_{\nu }
-\partial _{\nu }A_{\mu }$ stands for the usual field strength tensor of the
$U(1)$ gauge field $A_{\mu }$, which is minimally coupled to the $CP(2)$
sector via the covariant derivative $\nabla _{\mu }\phi =D_{\mu }\phi
-h^{-1}\left( \phi ^{\dag}D_{\mu }\phi \right) \phi $. Here, $D_{\mu }\phi $
is given by
\begin{equation}
D_{\mu }\phi =\partial _{\mu }\phi -igA_{\mu }\mathbb{Q}\phi \text{,}
\label{devcov1}
\end{equation}%
where $g$ represents an electromagnetic coupling constant and $\mathbb{Q}$
stands for a real charge matrix (diagonal and traceless). The topological
current density of the gauged $CP(2)$ reads
\begin{equation}
\hspace{-0.1cm}T_{\mu }=\frac{1}{2\pi ih}\varepsilon _{\mu \nu \rho }\left[
\left( \nabla ^{\nu }\phi \right) ^{\dag }\left( \nabla ^{\rho }\phi \right)
-\frac{ig}{2}F^{\nu \rho }\left( \phi ^{\dag }\mathbb{Q}\phi \right) \right]
\!\!\text{,}  \label{CPch03}
\end{equation}%
its topological charge being given by
\begin{equation}
\mathfrak{Q}=\int d^{2}\mathbf{x}\,T_{0}\in \mathbb{Z}\setminus \left\{
0\right\} \text{.}  \label{CPch04}
\end{equation}

As it was mentioned previously, the $CP(2)$ configuration (\ref{beta1}) {%
possesses} {a }null topological charge {in} both the free case (\ref{CPch02}%
) {and} the gauged case (\ref{CPch04}) where the charge matrix $\mathbb{Q}$
is given by \cite{loginov}
\begin{equation}
\mathbb{Q}=\frac{1}{2}\text{diag}\left( 1,-1,0\right) ,
\end{equation}%
which is related to the matrix $\lambda _{3}=\text{diag} \left(
1,-1,0\right) $, i.e., one of the Gell-Mann matrices which represent the $%
SU(3)$ group.

In the remaining of {present Section}, we will show {that} the effective
model for the configuration (\ref{beta1}) {obtained} from the original {%
theory} (\ref{1m0}) supports a well-defined BPS structure. Thus, the
Lagrangian density {which describes} the effective model is%
\begin{eqnarray}
\mathcal{L} &=&-\frac{1}{4}F_{\mu \nu }F^{\mu \nu }+\left( D_{\mu }\phi
\right) ^{\dag }D^{\mu }\phi  \notag \\[0.12in]
&&-U_{0}(\phi _{3})-\lambda \left( h-\phi ^{\dag }\phi \right) \text{,}
\label{1m0a}
\end{eqnarray}%
where $\phi $ {represents }the configuration (\ref{beta1}) {and} the
covariant derivative $D_{\mu }\phi $ {is} defined in {the} Eq.(\ref{devcov1}%
). {Also,}\ $\lambda ${\ stands for }a Lagrange multiplier {which guarantees}
the condition $h=\phi ^{\dag }\phi =2|\psi |^{2}+(\phi _{3})^{2}$.

The field {equation} for the gauge {sector reads}%
\begin{equation}
\partial _{\nu }F^{\nu \mu }=J^{\mu }\text{,}  \label{EQL01}
\end{equation}%
where $J^{\mu }$ is the conserved current density {related to} the charged
field $\psi ${, its expression being}%
\begin{equation}
J^{\mu }=ig\left[ \left( \hat{D}^{\mu }\psi \right) ^{\ast }\psi -\psi
^{\ast }\hat{D}^{\mu }\psi \right] \text{,}  \label{corrente}
\end{equation}%
{with} the quantity $\hat{D}^{\mu }\psi $ {standing for} the {corresponding}
covariant derivative{, i.e.}%
\begin{equation}
\hat{D}_{\mu }\psi =\partial _{\mu }\psi -\frac{ig}{2}A_{\mu }\psi \text{.}
\label{devcov1x}
\end{equation}

{On the other hand,} the field equation for the charged {sector }$\psi ${\
itself is}%
\begin{equation}
\hat{D}_{\mu }\hat{D}^{\mu }\psi -\lambda \psi =0\text{,}  \label{EQL02}
\end{equation}%
{while the one} for the neutral field $\phi _{3}$ {reads}%
\begin{equation}
\partial _{\mu }\partial ^{\mu }\phi _{3}+\frac{1}{2}\frac{\partial U_{0}}{%
\partial \phi _{3}}-\lambda \phi _{3}=0\text{.}  \label{EQL03}
\end{equation}

{Via the combination between }the last two {equations} and the relation $%
h=2\left\vert \psi \right\vert ^{2}+\left( \phi _{3}\right) ^{2}$, we {%
additionally} attain the following expression for the Lagrange multiplier $%
\lambda $,{\ i.e.}%
\begin{equation}
h\lambda =-2\left\vert \hat{D}_{\mu }\psi \right\vert ^{2}-\left( \partial
_{\mu }\phi _{3}\right) ^{2}+\frac{1}{2}\phi _{3}\frac{\partial U_{0}}{%
\partial \phi _{3}}\text{.}  \label{EQL04}
\end{equation}

{We now} write {down} the {equations for} stationary {fields}. {In this
sense, the} Eq. (\ref{EQL01}) {leads to} the Gauss law%
\begin{equation}
\partial _{k}\partial _{k}A_{0}=g^{2}A_{0}|\psi |^{2}\text{,}  \label{EQL05}
\end{equation}%
{which} is identically satisfied by the gauge condition $A^{0}=0$. {This
condition therefore stands for the gauge choice which} we use along {the
rest of} this {Section}. Hence, we conclude that the stationary solutions
inherent to the model (\ref{1m0a}) present zero total charge and carry\ {only%
} magnetic flux.

{The }Amp\`{e}re's law becomes%
\begin{equation}
\epsilon _{kj}\partial _{j}B=-J_{k}\text{,}  \label{EQL06}
\end{equation}%
{while the} stationary equations for the fields $\psi $ and $\phi _{3} $ are%
\begin{eqnarray}
&&\displaystyle\hat{D}_{k}\hat{D}_{k}\psi +\lambda \psi =0\text{,}
\label{EQL07} \\[0.2cm]
&&\displaystyle\partial _{k}\partial _{k}\phi _{3}-\frac{1}{2}\frac{\partial
U_{0}}{\partial \phi _{3}}+\lambda \phi _{3}=0\text{,}  \label{EQL08}
\end{eqnarray}%
with $\lambda $ now written as%
\begin{equation}
h\lambda =2\left\vert \hat{D}_{k}\psi \right\vert ^{2}+\left( \partial
_{k}\phi _{3}\right) ^{2}+\frac{1}{2}\phi _{3}\frac{\partial U_{0}}{\partial
\phi _{3}}\text{.}  \label{EQL09}
\end{equation}

\subsection{The BPS structure of the model described by the configuration $(%
\protect\ref{beta1})$}

The stationary energy density of the model (\ref{1m0a}) is%
\begin{equation}
\varepsilon =\frac{1}{2}B^{2}+\left( D_{k}\phi \right) ^{\dag }D_{k}\phi
+U_{0}(\phi _{3})\text{,}
\end{equation}%
where {we have used }the gauge condition $A_{0}=0$. {The} corresponding
total energy is given by%
\begin{equation}
\mathcal{E}=\int d^{2}\mathbf{x~}\varepsilon \text{.}
\end{equation}

{In order to} implement the BPS formalism, we consider the relations%
\begin{equation}
\frac{1}{2}B^{2}=\frac{1}{2}\left( B\mp \sqrt{2U_{0}}\right) ^{2}\pm B\sqrt{%
2U_{0}}\text{,}
\end{equation}%
and
\begin{eqnarray}
\left\vert D_{k}\phi \right\vert ^{2} &=&\frac{1}{2}\left\vert D_{j}\phi \pm
ih^{-1/2}\epsilon _{jk}\left( \phi \times D_{k}\phi \right) ^{\ast
}\right\vert ^{2}  \notag \\[0.2cm]
&&\mp ih^{-1/2}\epsilon _{jk}\phi \cdot \left( D_{j}\phi \times D_{k}\phi
\right) \text{,}
\end{eqnarray}%
{via which we rewrite the total energy in the form}%
\begin{eqnarray}
\mathcal{E} &=&\int d^{2}\mathbf{x}\left\{ \frac{1}{2}\left\vert D_{j}\phi
\pm ih^{-1/2}\epsilon _{jk}\left( \phi \times D_{k}\phi \right) ^{\ast
}\right\vert ^{2}\right.  \notag \\[0.2cm]
&&+\frac{1}{2}\left( B\mp \sqrt{2U_{0}}\right) ^{2}\pm B\left( \sqrt{2U_{0}}%
-h^{1/2}g\phi _{3}\right)  \notag \\[0.2cm]
&&\left. \mp \frac{i}{h^{1/2}}\left[ \epsilon _{jk}\phi \cdot \left(
D_{j}\phi \times D_{k}\phi \right) +ihg\phi _{3}B\right] \right\} \text{.}%
\quad  \label{Ebps1}
\end{eqnarray}

{Now, whether we consider} $B=-F_{12}=-\epsilon _{jk}\partial _{j}A_{k}$,
the last term {in the Equation above} {can be} related to $0^{\text{th}}$%
-component of the topological current density of the model, {i.e.}%
\begin{equation}
\bar{q}_{\mu }=\frac{\epsilon _{\mu \nu \lambda }}{i2\pi h^{3/2}}\left[ \phi
\cdot \left( D^{\nu }\phi \times D^{\lambda }\phi \right) -ih\frac{g}{2}\phi
_{3}F^{\nu \lambda }\right] \text{,}  \label{TPccc}
\end{equation}%
whose integration provides the topological charge%
\begin{equation}
\int d^{2}\mathbf{x~}\bar{q}_{0}\in \mathbb{Z}\setminus \left\{ 0\right\}
\text{,}  \label{TPcharge}
\end{equation}%
{from which we conclude that }the integration of the term in the third row
of {the} Eq. (\ref{Ebps1}) provides\ the {system's} BPS energy in terms of
the topological charge, {i.e.}%
\begin{equation}
\mathcal{E}_{bps}=\pm 2\pi h\int d^{2}\mathbf{x~}\bar{q}_{0}>0\text{.}
\label{ebpssf}
\end{equation}

{In the} second row of {the} Eq. (\ref{Ebps1}), we set the factor {which
multiplies} the magnetic field {as being zero}, {from which} we obtain the
BPS\ potential%
\begin{equation}
U_{0}=\frac{1}{2}hg^{2}(\phi _{3})^{2}\text{. }
\end{equation}

{In view of} the two last {equations},\ the total energy {then} becomes%
\begin{eqnarray}
\mathcal{E} &=&\mathcal{E}_{bps}+\frac{1}{2}\int d^{2}\mathbf{x}\left( B\mp
h^{1/2}g\phi _{3}\right) ^{2}  \notag \\[0.2cm]
&&+\frac{1}{2}\int d^{2}\mathbf{x~}\left\vert D_{j}\phi \pm
ih^{-1/2}\epsilon _{jk}\left( \phi \times D_{k}\phi \right) ^{\ast
}\right\vert ^{2}\text{,}
\end{eqnarray}%
{from which} we see that total energy {satisfies}%
\begin{equation}
\mathcal{E}\geq \mathcal{E}_{bps}\text{,}
\end{equation}%
with the equality being satisfied {when} the quadratic terms\ {which appear}
{within }the integrals {are chosen as being zero}. This {choice }provides
the BPS or self-dual equations of the system, i.e.%
\begin{equation}
B=\pm h^{1/2}g\phi _{3}\text{,}  \label{fbps1}
\end{equation}%
\begin{equation}
D_{j}\phi =\mp ih^{-1/2}\epsilon _{jk}\left( \phi \times D_{k}\phi \right)
^{\ast }\text{,}  \label{fbps2}
\end{equation}%
{which resemble} the ones {obtained in the context of} the gauged $O(3)$
sigma model. Indeed, the BPS configurations can be considered as {the}
classical solutions related to an extended supersymmetric {version} \cite%
{witten,spector} of the model (\ref{1m0a}).

{In particular,} the second BPS equation {can be rewritten} in terms of the
field components {which appear} in (\ref{beta1}). {We then get}%
\begin{eqnarray}
\hat{D}_{j}\psi &=&\mp ih^{-1/2}\epsilon _{jk}\left( \psi \partial _{k}\phi
_{3}-\phi _{3}\hat{D}_{k}\psi \right) \text{,}  \label{fbps3} \\[0.2cm]
\partial _{j}\phi _{3} &=&\pm h^{-1/2}g^{-1}\epsilon _{jk}J_{k}\text{,}
\label{fbps4}
\end{eqnarray}%
where $J_{k}$ is the conserved current density defined previously in {the}
Eq. (\ref{corrente}).

In the BPS limit, the self-dual equations recover the stationary Amp\`{e}%
re's law (\ref{EQL06}) and the stationary Euler-Lagrange equations for the
fields $\psi $ (\ref{EQL07}) and $\phi _{3}$\ (\ref{EQL08}).

On the other hand, the solutions of the BPS equations describing radially
symmetric vortices were recently studied in the refs. \cite{loginov,casana},
for the case $\beta =\beta _{1}$.

\section{Maxwell-$CP(2)$ vortex-like solitons in the presence of a magnetic
impurity \label{sec2}}

\label{general}

We begin this work defining the first model we will investigate. It consists
of a Maxwell-$CP(2)$ theory extended to include an additional term
representing the presence of a magnetic impurity. The resulting Lagrange
density describing the enlarged model is
\begin{equation}
\mathcal{L}=-\frac{1}{4}F_{\mu \nu }F^{\mu \nu }+\left( D_{\mu }\phi \right)
^{\dag }\left( D^{\mu }\phi \right) -U\left( \phi _{3},\Delta \right)
+\Delta B\text{,}  \label{1m}
\end{equation}%
where $\phi $ {stands for} the $CP(2)$ field configuration defined in {the}
Eq. (\ref{beta1}).

The third term in {the} Eq. (\ref{1m}) is the potential $U=U(\phi
_{3},\Delta )$ {which} also depends on the function $\Delta $ (the so-called
\textit{magnetic impurity}). The last term couples the magnetic field $B$ to
the impurity $\Delta $ {which}, in our analysis, depend explicitly on the
spatial coordinates (i.e. $\Delta =\Delta (|\mathbf{x}|)$) and therefore
breaks the translational invariance of the model. {This} breaking is not a
problem whether we consider the model (\ref{1m}) as an effective one. The
point here is that the function $\Delta $ represents a magnetic impurity in {%
a} medium where vortices exist, see the arguments in \cite{15}.

The {equation for the} gauge field is given by
\begin{equation}
\partial _{\nu }F^{\nu \mu }+\left( \delta _{2}^{\mu }\partial _{1}\Delta
-\delta _{1}^{\mu }\partial _{2}\Delta \right) =J^{\mu }\text{,}
\end{equation}%
where $J^{\mu }$ is the current density (\ref{corrente}).

We highlight {that} the presence of the term $\Delta B$ in (\ref{1m}) does
not change the structure of the Gauss law obtained in the context of the
usual Maxwell-$CP(2)$ model (\ref{1m0a})\ \textit{without} {the} magnetic
impurity, {see the} Ref. \cite{casana}. {In this sense}, the stationary
Gauss law is {still} given by {the} Eq. (\ref{EQL05}). {As a consequence,}
we conclude that the stationary solutions inherent to the model (\ref{1m})
also present {no electric} charge and only carry magnetic flux.

We {thus} focus our attention on those time-independent configurations {with
rotational symmetry} {which} transport only magnetic flux, {from which} we
use the map%
\begin{eqnarray}
&\displaystyle A_{i}=-\frac{\epsilon _{ij}x_{j}}{gr^{2}}A(r)\text{,}&
\label{2m} \\
&\displaystyle\psi =\sqrt{\frac{h}{2}}\,e^{im\theta }\sin \alpha (r)\text{ \
\ and \ \ }\phi _{3}=\sqrt{h}\cos \alpha (r)\text{,}&\quad  \label{3m}
\end{eqnarray}%
where $\epsilon _{ij}$ stands for the two-dimensional Levi-Civita's symbol
(with $\epsilon _{12}=+1$), $r$ and $\theta $ represent the polar
coordinates, {while} $m\in \mathbb{Z}\setminus \left\{ 0\right\}$ is the
winding number of the {resulting} configuration.

Under the parametrization (\ref{2m}) the magnetic field reads%
\begin{equation}
B(r)=-\frac{1}{gr}\frac{dA}{dr}\text{.}  \label{x7}
\end{equation}

{The both profile functions $A(r)$ and $\alpha (r)$ depending only on the
radial coordinate} must describe regular configurations with finite energy,
from which they are supposed to satisfy the usual boundary conditions, i.e.,
\begin{eqnarray}
& \displaystyle \alpha (r=0)=0\text{ \ and \ }A(r=0)=0\text{,} &  \label{31m}
\\[0.2cm]
& \displaystyle\alpha \left( r\rightarrow \infty \right) \rightarrow \frac{%
\pi }{2}\text{ \ \ and \ \ }A\left( r\rightarrow \infty \right) \rightarrow
2m\text{.}&  \label{bc2m}
\end{eqnarray}

We now look for the first-order framework inherent to the model {(\ref{1m})
through the} standard BPS prescription, i.e., via the minimization of the {%
enlarged model's total energy. The} starting point is the expression for the
corresponding energy distribution. {In this sense, given the} rotationally
symmetric map (\ref{2m}) and (\ref{3m}) and {all the} conventions introduced
above, the time-independent energy density can be {written in the form}%
\begin{eqnarray}
\varepsilon &=&\frac{1}{2}B^{2}+U(\alpha ,\Delta )-\Delta B  \notag \\[0.2cm]
&&+h\left( \frac{d\alpha }{dr}\right) ^{2}+h\frac{\left( 2m-A\right) ^{2}}{%
4r^{2}}\sin ^{2}\alpha \text{,}  \label{edm}
\end{eqnarray}%
from which one gets the total energy $\mathcal{E}$ as%
\begin{eqnarray}
\frac{\mathcal{E}}{2\pi } &=&\int_{0}^{\infty }\left[ \frac{1}{2}%
B^{2}+U(\alpha ,\Delta )-\Delta B\right.  \notag \\[0.2cm]
&&\left. +h\left( \frac{d\alpha }{dr}\right) ^{2}+h\frac{\left( 2m-A\right)
^{2}}{4r^{2}}\sin ^{2}\alpha \right] rdr\text{.}\quad \quad
\end{eqnarray}

{After some algebra, the implementation of the BPS formalism leads to the
following expression for the total energy:
\begin{eqnarray}
\frac{\mathcal{E}}{2\pi } &=&\int_{0}^{\infty }\left[ h\left( \frac{d\alpha
}{dr}\pm \frac{(2m-A)}{2r}\sin \alpha \right) ^{2}\right.   \notag \\[0.2cm]
&&+\frac{1}{2}\left( B\mp \sqrt{2U}\right) ^{2}\pm 2\pi h\bar{q}_{0}  \notag
\\[0.2cm]
&&\left. \pm B\left( \sqrt{2U}-gh\cos \alpha \mp \Delta \right) \!\!\frac{{}%
}{{}}\right] rdr\text{,}\hspace{1cm}  \label{5mxx}
\end{eqnarray}%
in which we have used the expression (\ref{x7}) for the magnetic field to
attain the third and fourth terms. The quantity $\bar{q}_{0}$ is the
topological charge density defined from Eq. (\ref{TPccc}), which expressed
in polar coordinates reads
\begin{equation}
\bar{q}_{0}=\frac{1}{2\pi r}\frac{d}{dr}\left[ (2m-A)\cos \alpha \right] .
\label{qq0}
\end{equation}%
}

To complete the implementation of the BPS prescription, we set to zero the
expression multiplying the magnetic field in the third row of the Eq. (\ref%
{5mxx}). It fixes the BPS potential of the enlarged model, in terms of both
the profile $\alpha (r)$ and the impurity $\Delta (r)$ itself, as
\begin{equation}
U(\alpha ,\Delta )=\frac{g^{2}h^{2}}{2}\left( \cos \alpha \pm \frac{\Delta }{%
gh}\right) ^{2}\text{,}  \label{x14a}
\end{equation}%
where both the potential and the function $\Delta $ vanish when $%
r\rightarrow \infty $.

{Thus, by considering the relation (\ref{x14a}), we then write the total
energy (\ref{5mxx}) as}%
\begin{eqnarray}
\frac{\mathcal{E}}{2\pi } &=&\frac{\mathcal{E}_{bps}}{2\pi }+\frac{1}{2}%
\int_{0}^{\infty }\left[ B\mp gh\left( \cos \alpha \mp \frac{\Delta }{gh}%
\right) \right] ^{2}rdr  \notag \\[0.2cm]
&&+h\int_{0}^{\infty }\left[ \frac{d\alpha }{dr}\pm \frac{(2m-A)}{2r}\sin
\alpha \right] ^{2}rdr\text{,}  \label{x5}
\end{eqnarray}%
where $\mathcal{E}_{bps}$ is defined by the Eq. (\ref{ebpssf}) with $\bar{q}%
_{0}$ given in (\ref{qq0}). So, the quantity $\mathcal{E}_{bps}$ stands for
the lower bound (i.e., the Bogomol'nyi bound) of the total energy of the
rotationally symmetric configurations. The particular value of the
Bogomol'nyi bound can be calculated by using the boundary conditions (\ref%
{31m}) and (\ref{bc2m}). Therefore, the BPS energy for the model (\ref{1m})
becomes
\begin{equation}
\mathcal{E}_{bps}=\mp 4\pi hm >0\text{,}  \label{8m}
\end{equation}%
where the upper (lower) sign holds for negative (positive) values of the
winding number $m$.

Therefore, from the Eq. (\ref{x5}), it is possible to note that the total
energy of the system satisfies the inequality
\begin{equation}
\mathcal{E}\geq \mathcal{E}_{bps}\text{,}
\end{equation}%
{with the lower bound attained when the fields satisfy the so called BPS
equations, i.e.}%
\begin{eqnarray}
&\displaystyle B=\pm gh\left( \cos \alpha \pm \frac{\Delta }{gh}\right)
\text{,} &  \label{bps1m} \\[0.2cm]
&\displaystyle\frac{d\alpha }{dr}=\mp \frac{(2m-A)}{2r}\sin \alpha \text{,} &
\label{bps2m}
\end{eqnarray}
whose solutions are rotationally symmetric structures with total energy
equal to
\begin{equation}
\mathcal{E}=\mathcal{E}_{bps}=4\pi h\left\vert m\right\vert \text{.}
\label{xm}
\end{equation}

Note that the value in (\ref{xm}) is quantized according to the winding
number $m$, such as expected for topological configurations. We also
highlight that the Bogomol'nyi bound is not affected by the presence of the
magnetic impurity.

In the BPS limit, the energy density (\ref{edm}) can be rewritten in the form%
\begin{equation}
\varepsilon _{bps}=\left( \sqrt{2U}\mp \frac{\Delta }{2}\right) ^{2}-\frac{%
\Delta ^{2}}{4}+2h\left( \frac{d\alpha }{dr}\right) ^{2}\text{,}
\label{EEbps}
\end{equation}%
where $U$ stands for the BPS potential in the Eq. (\ref{x14a}).

The potential can also be written as a function of $\phi _{3}$ and $\Delta$,
\begin{equation}
U\left( \phi _{3},\Delta \right) =\frac{g^{2}h}{2}\left( \phi _{3}\pm \frac{%
\Delta }{g\sqrt{h}}\right) ^{2}\text{,}  \label{x14c}
\end{equation}%
which spontaneously breaks the $SU(3)$ symmetry inherent to the model (\ref%
{1m}), as expected. Also, the expression in (\ref{x14c}) reveals that the
presence of the magnetic impurity in the original Lagrangian density (\ref%
{1m}) requires an adjustment on the potential (in comparison to the model
\textit{without} the impurity) to support the existence of first-order
configurations.

In the next Section, we consider a localized magnetic impurity of the
Gaussian-type, from which we solve the BPS equations (\ref{bps1m}) and (\ref%
{bps2m}) numerically by means of a finite-difference scheme according the
boundary conditions (\ref{31m}) and (\ref{bc2m}).


\subsection{The Maxwell-$CP(2)$ vortex-like solitons: numerical analysis}

In order to continue, we need to choose an explicit expression for the
localized magnetic impurity. For the sake of convenience, we prefer to work
with a Gaussian profile centered at the origin, i.e.%
\begin{equation}
\Delta (r)=c e^{-dr^{2}} \text{,}  \label{mim}
\end{equation}%
{where both $c$ and $d\in \mathbb{R}$}, with {$d>0$.} In this case, the
parameters $c$ and $d$ control the height and width of the impurity,
respectively.

Now, given the impurity (\ref{mim}), the first-order potential (\ref{x14a})
can be written as%
\begin{equation}
U=\frac{g^{2}h^{2}}{2}\left( \cos \alpha \pm \frac{c}{gh} e^{-dr^{2}}
\right) ^{2}\text{,}  \label{xixim}
\end{equation}%
in which the upper (lower) sign holds for negative (positive) values of $m$ {%
(the winding number). In this manuscript, }we consider the case with $d=1$
and different values for $c$ {given that, as we explain later below, this
case gives} rise to interesting modifications on the profiles of the
resulting first-order solutions.

\begin{figure}[tbp]
\includegraphics[width=8.4cm]{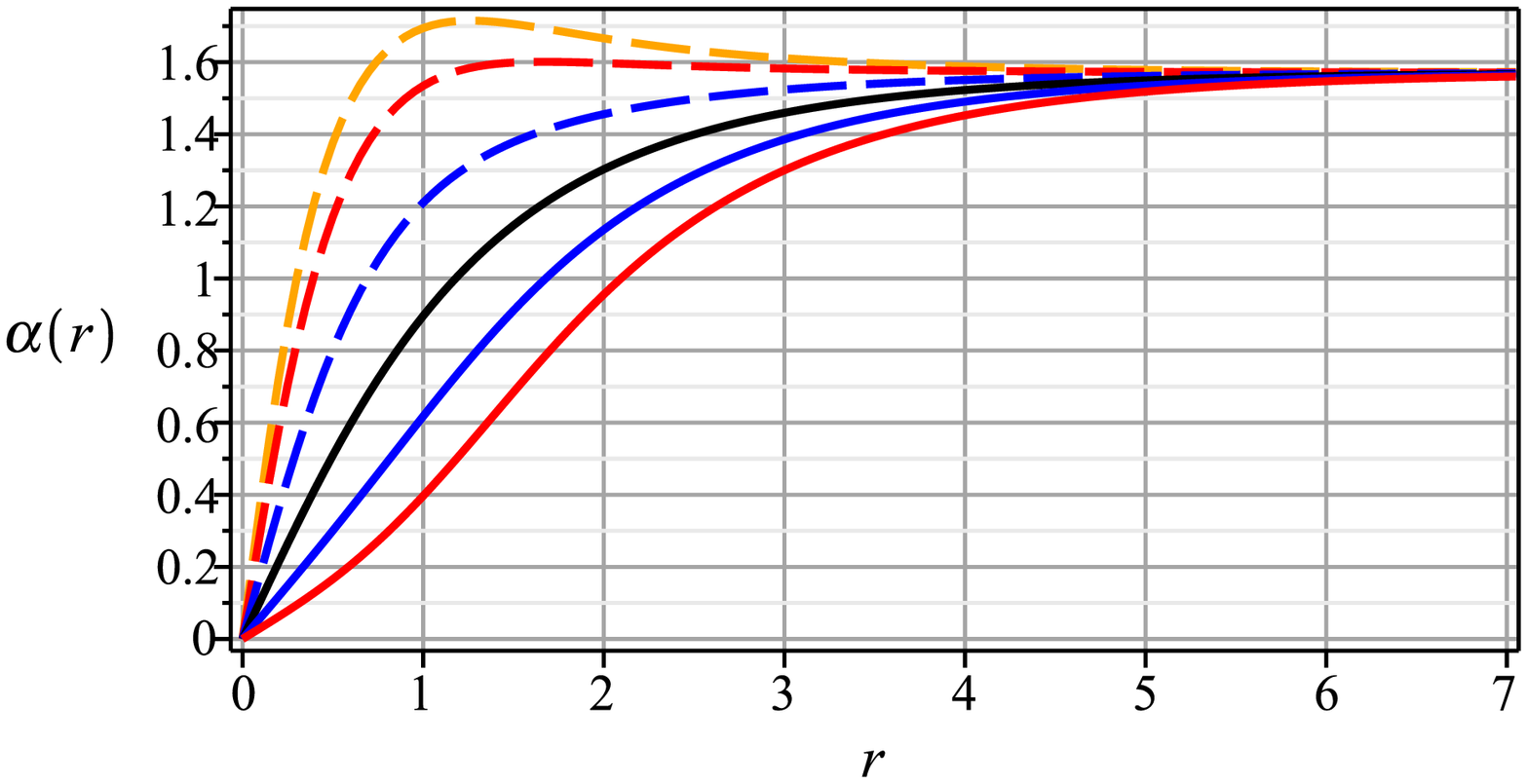} %
\includegraphics[width=8.4cm]{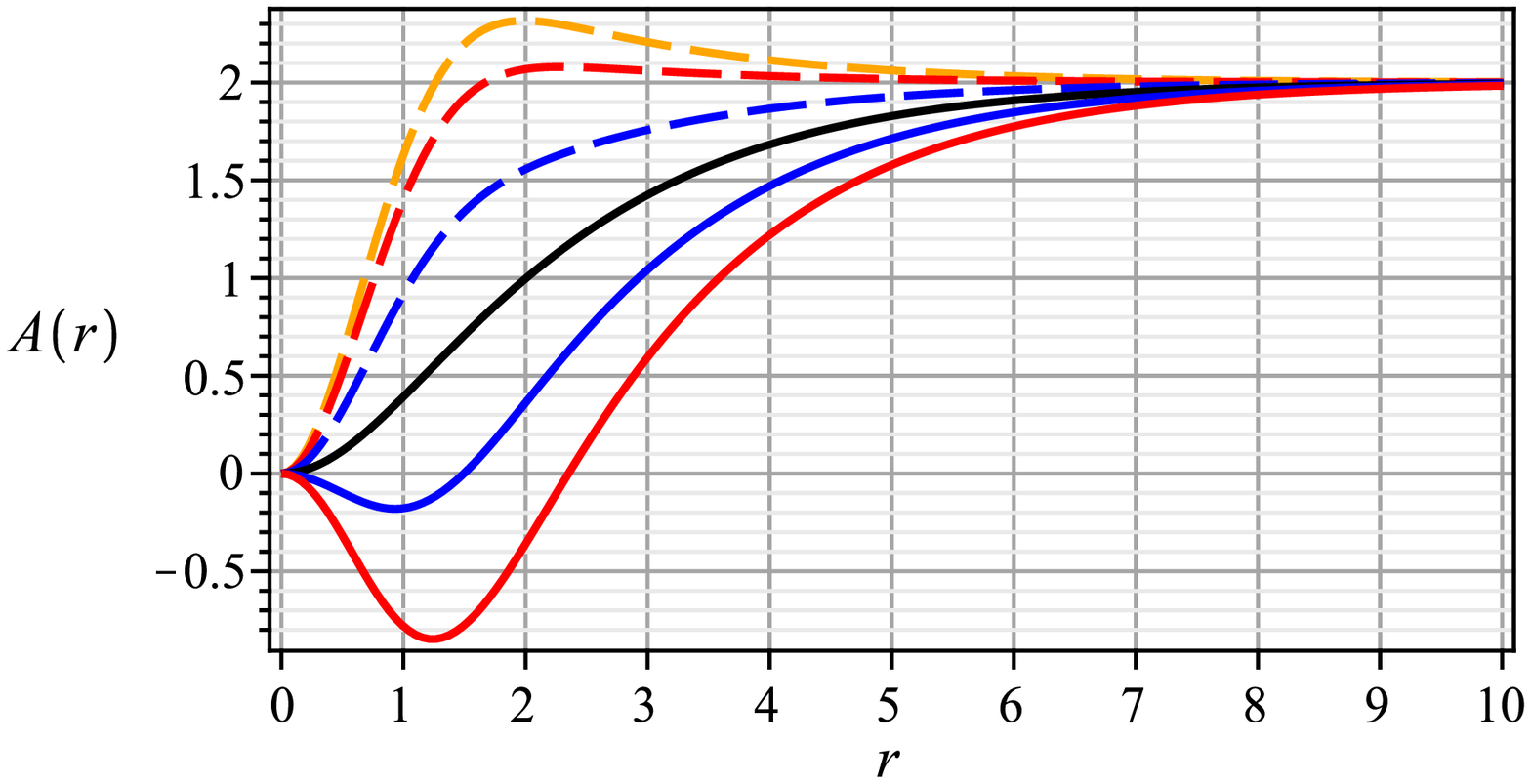}
\caption{Numerical solutions to $\protect\alpha (r)$ (top) and $A(r)$
(bottom) coming from (\protect\ref{bpsx}) and (\protect\ref{bpsy1}) in the
presence of (\protect\ref{31m}) and (\protect\ref{bc2m}). Here, we have
fixed $g=h=1$, $m=1$ (lower signs in the first-order expressions) and $d=1$.
This Figure shows the profiles for $c=-5$ (dashed orange line), $c=-4$
(dashed red line), $c=-2$ (dashed blue line), $c=0$ (usual solution, no
impurity, solid black line), $c=+2$ (solid blue line), $c=+4$ (solid red
line).}
\label{figg1}
\end{figure}

In view of the potential (\ref{xixim}), the first-order equations (\ref%
{bps1m}) and (\ref{bps2m}) {take the form}%
\begin{eqnarray}
&\displaystyle\frac{1}{r}\frac{dA}{dr}=\mp g^{2}h\left( \cos \alpha \pm
\frac{c}{gh}e^{-dr^{2}}\right) \text{,}&  \label{bpsx} \\[0.08in]
&\displaystyle\frac{d\alpha }{dr}=\mp \frac{(2m-A)}{2r}\sin \alpha \text{,}&
\label{bpsy1}
\end{eqnarray}%
which must be solved according the boundary conditions (\ref{31m}) and (\ref%
{bc2m}). Here, we have also used the Eq. (\ref{x7}) for the magnetic field.

In order to solve the above first-order equations numerically, {we fix $%
g=h=1 $, for simplicity. Furthermore,} we choose $m=1$ (i.e. lower signs in
the first-order expressions) and $d=1$ (a fixed value for the width of the
impurity). We then study the resulting first-order equations through a
finite-difference algorithm for different values of $c$ (the height of the
impurity). {Subsequently, we depict the numerical profiles for the profile
functions $\alpha (r)$ and $A(r)$,} the magnetic field $B(r)$ and the energy
density $\varepsilon _{bps}(r)$.

The Figure \ref{figg1} {shows the field profiles} $\alpha (r)$ and $A(r)$
for $c=-5$ (dashed orange line), $c=-4$ (dashed red line), $c=-2$ (dashed
blue line), $c=0$ (usual solution, no impurity, solid black line), $c=+2$
(solid blue line) and $c=+4$ (solid red line). {It is important to point
that the field profiles lose the original monotonicity (attained in the
\textit{absence} of impurity) as} the value of $|c|$ increases. As a
consequence, the profiles for $\alpha $ and $A$ present a global maximum for
$c=-4$ and $c=-5$, while $A$ presents a global minimum for $c=2$ and $c=4$.

\begin{figure}[tbp]
\includegraphics[width=8.4cm]{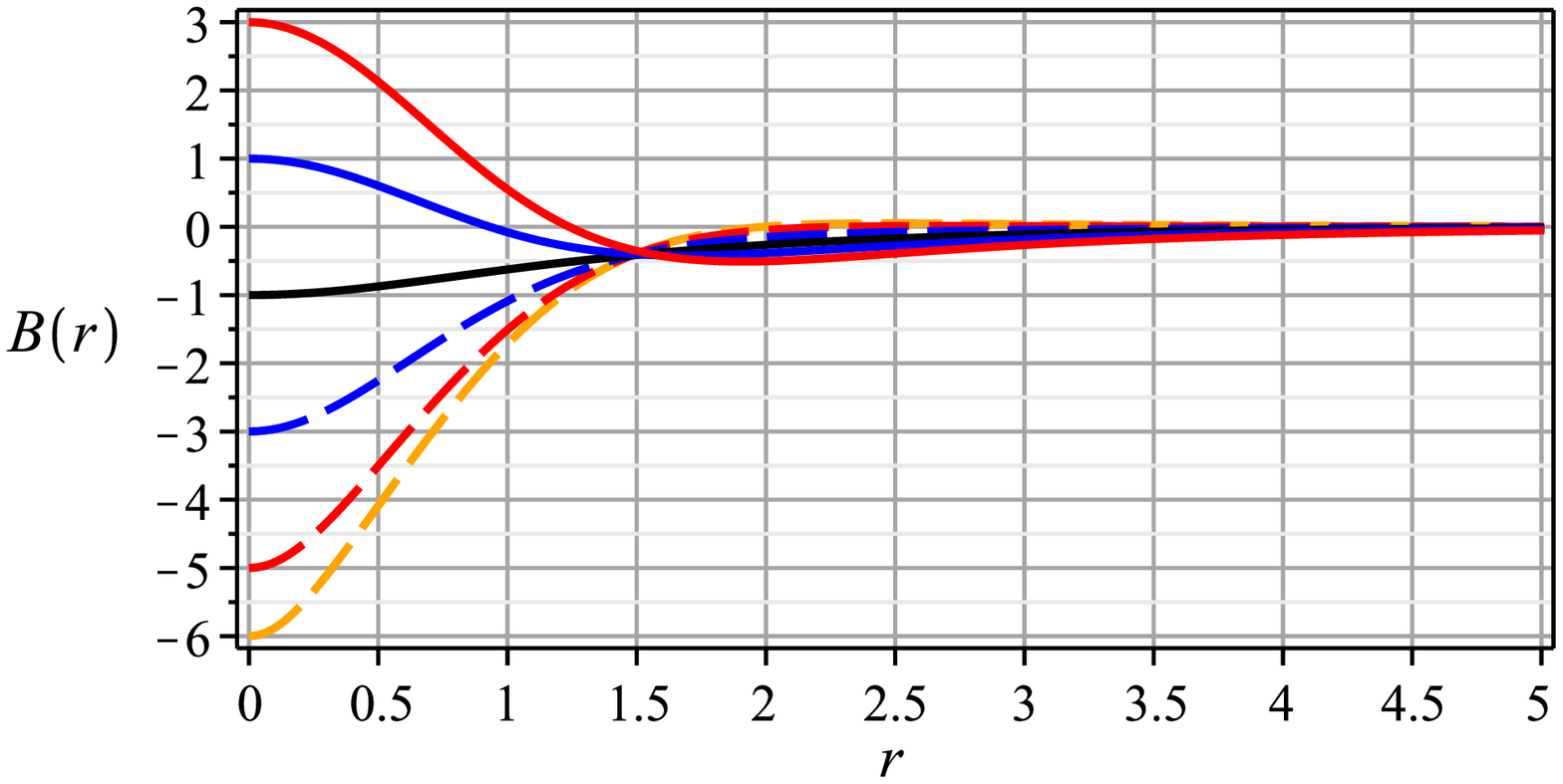} %
\includegraphics[width=8.4cm]{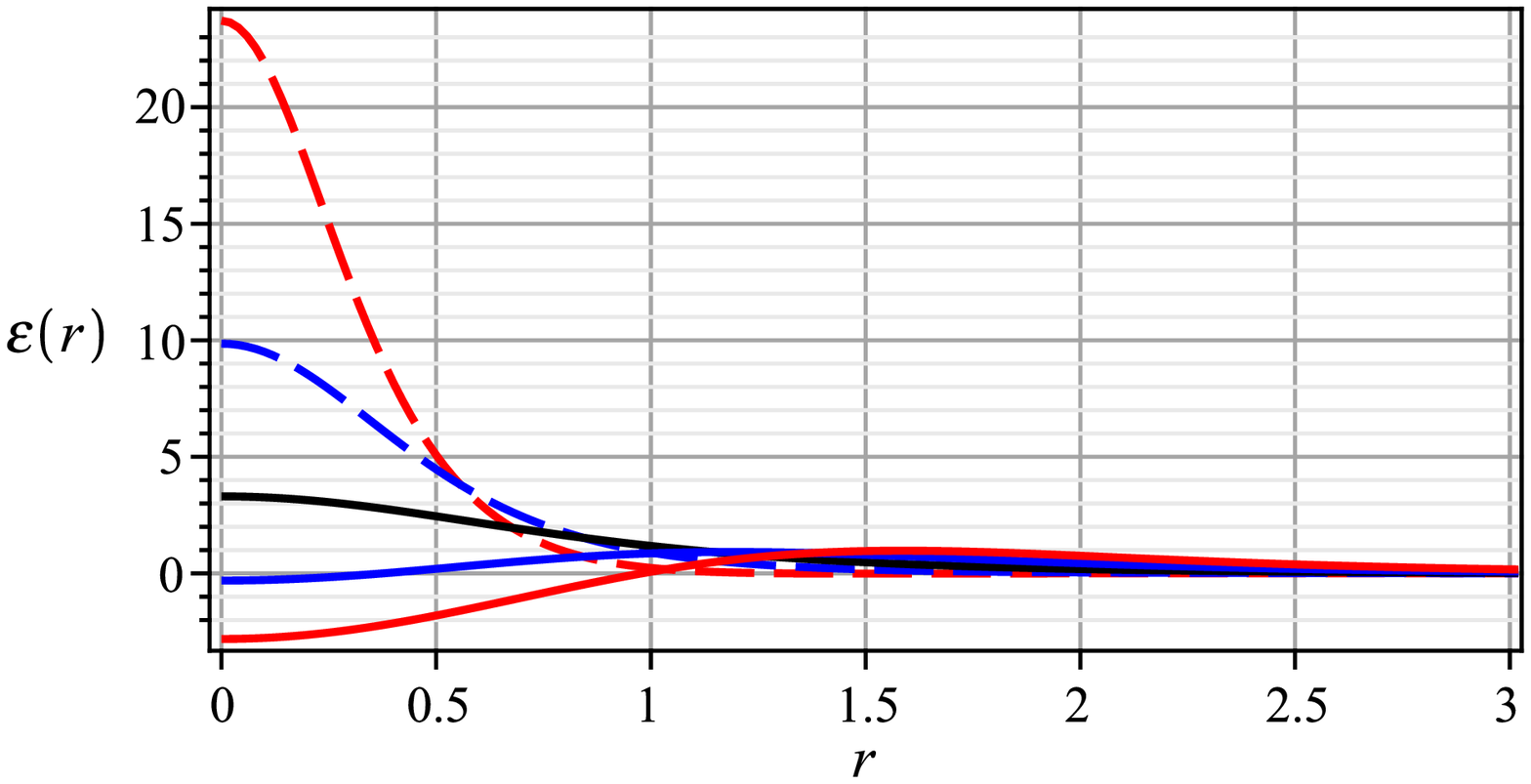}
\caption{Numerical solutions to the magnetic field $B(r)$ (top) and the
energy density $\protect\varepsilon (r)$ (bottom) of the first-order BPS
Maxwell-$CP(2)$ configurations. Conventions as in the Fig. \protect\ref%
{figg1}.}
\label{figg2}
\end{figure}

In the Figure \ref{figg2}, we depict the solutions {for both} the magnetic
field $B(r)$ and the BPS energy density $\varepsilon _{bps}$. {The profiles}
suggest that the parameter $c$ (the height of the impurity) induces an
inversion on the sign of both the magnetic field and the BPS energy density
as these solutions approach the origin. {In particular, for $c>0$, the BPS
energy density reaches negative values within a finite spatial region
beginning at $r=0$. The negative values arise due to the magnetic impurity
which precludes to express the BPS energy density (\ref{edm}) as a sum of
positive terms only, see the Eq. (\ref{EEbps}).}

\subsubsection{Behavior of the solutions near the origin}

To explain the sign inversion of the magnetic field and the BPS energy
density near the origin, we first study the behavior of the profile
functions $\alpha (r)$ and $A(r)$ themselves. In this sense, for $m>0$, the
field profiles when $r\rightarrow 0$ behave as (here, $C_{0}>0\in \mathbb{R}$%
)
\begin{equation}
\alpha (r)\approx C_{0}r^{m}  \label{xa1}
\end{equation}%
\begin{equation}
A(r)\approx \frac{g^{2}h}{2}\left( 1-\frac{c}{gh}\right) r^{2}\text{,}
\label{xa2}
\end{equation}%
which promptly recover the usual results for $c=0$ (i.e. in the absence of
the magnetic impurity).

We see that the impurity does not change the behavior of the scalar profile
function $\alpha (r)$ near the origin. However, the impurity (via its height
parameter $c$) changes the factor which multiplies the relevant term in the
approximate solution for the gauge profile function $A(r)$.

{In order to present the behavior of both the magnetic field and the BPS
energy density near the origin, we consider $g=h=1$, $m=1$, and $d=1$ (i.e.
the values of the parameters used to obtain the numerical solutions). Then,
we get the following behavior for the magnetic field:%
\begin{equation}
B(r)\approx c-1+\frac{C_{0}^{2}-2c}{2}{r}^{2}\text{,}
\end{equation}%
and for the BPS energy density, we obtain%
\begin{equation}
\varepsilon _{bps}(r)\approx -c+1+2C_{0}^{2}+\frac{C_{1}}{6}{r}^{2}\text{,}
\end{equation}%
where $C_{1}=3c(3C_{0}^{2}+2)-4C_{0}^{2}(C_{0}^{2}+3)$. We observe that both
these expressions reflect the behavior presented in the Fig. \ref{figg2},
for }${r\rightarrow 0}${.}

\begin{figure}[tbp]
\includegraphics[width=8.4cm]{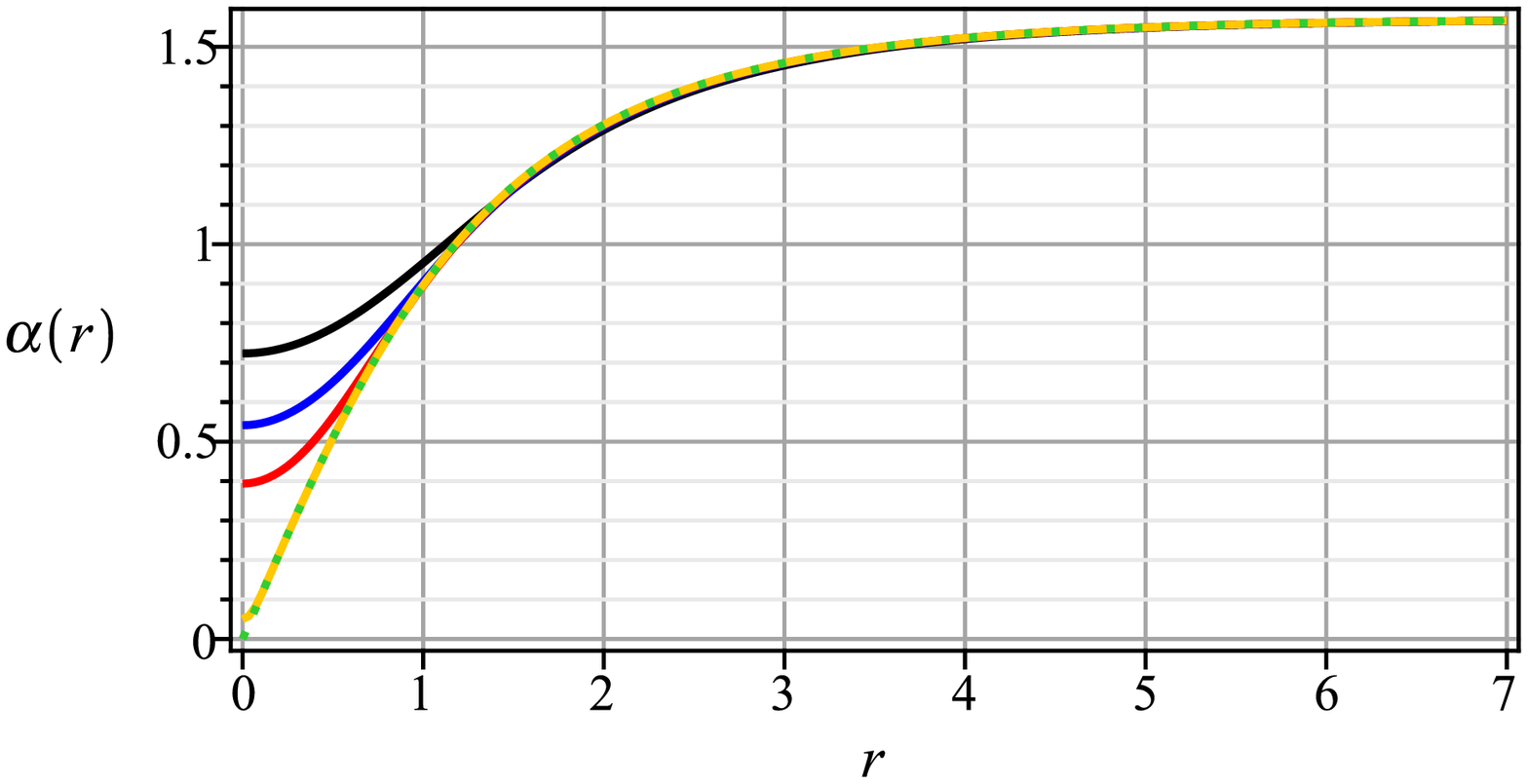} %
\includegraphics[width=8.4cm]{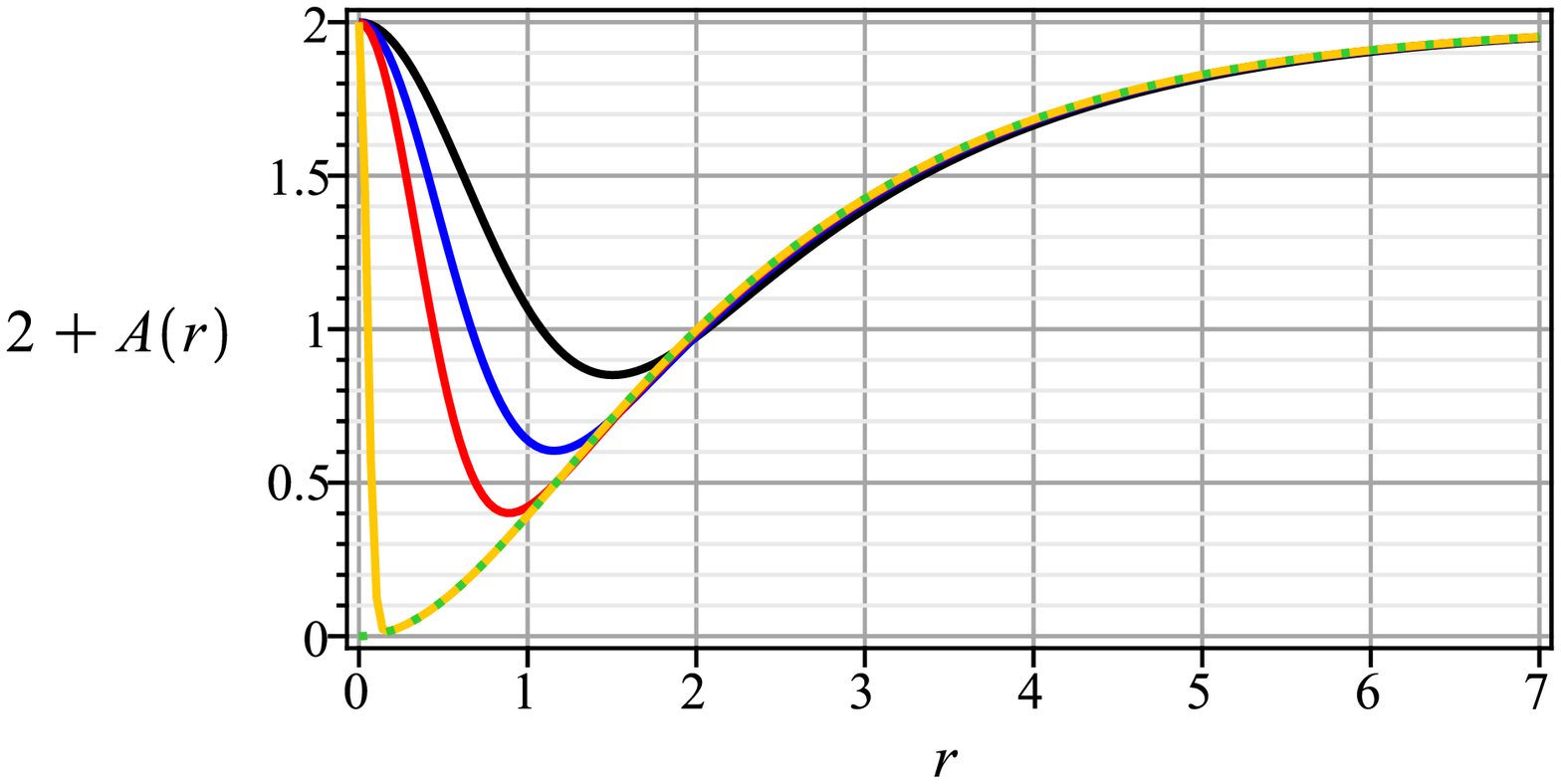}
\caption{Numerical solutions to $\protect\alpha (r)$ (top) and $2+A(r)$
(bottom) for $m=0$ (i.e. the topologically trivial configuration). Here, we
have depicted the solutions for $d=1$ ($c=4$, black line), $d=2$ ($c=8$,
blue line), $d=4$ ($c=16$, red line) and $d=256$ ($c=1024$, orange line).
The dotted green line represents the topological profile for $m=1$\ in the
absence of the impurity.}
\label{figg3x}
\end{figure}

\subsubsection{Behavior of the solutions in the asymptotic limit}

We also present the behavior of the profile fields $\alpha (r)$ and $A(r)$\
for large values of the radial {coordinate. In the present case, for all
values of $c$ and $d$ in (\ref{mim}), we have found that behavior of the
field profiles are}
\begin{eqnarray}
\displaystyle\alpha (r) &\approx &\frac{\pi }{2}-C_{\infty }\frac{\exp
\left( -Mr\right) }{\sqrt{r}}\text{,}  \label{ass1} \\[0.2cm]
\displaystyle A(r) &\approx &2m-2MC_{\infty }\sqrt{r}\exp \left( -Mr\right)
\text{,}  \label{ass2}
\end{eqnarray}%
{where $C_{\infty }$ stands for a positive real constant and}%
\begin{equation}
M=g\sqrt{\frac{h}{2}}\text{,}
\end{equation}%
which is the mass of both the scalar and gauge bosons. We then conclude that
the bosonic fields acquire the same mass in the self-dual limit, as in the
Maxwell-Higgs model.

Therefore, the expressions (\ref{ass1}) and (\ref{ass2}) reveal that, in the
presence of a localized magnetic impurity, the resulting first-order
vortices mimic the standard asymptotic behavior, i.e., a localized impurity
does not change the way the fields behave in the asymptotic region.

\subsection{The topologically trivial solution: the $m=0$ case \label{sec2b}}

\begin{figure}[t]
\includegraphics[width=8.4cm]{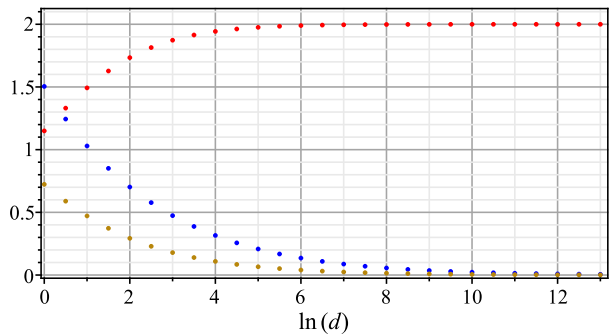}
\caption{The Maxwell-$CP(2)$ case with $m=0$: this Figure shows the behavior
of $-A(r_{min})$ (dotted red line), $r_{min}$ (dotted blue line) and $%
\protect\alpha (r=0)$ (dotted golden line) \textit{versus} ln$(d)$. Here, $%
r_{min}$ stands for the value of the radial coordinate for which the gauge
profile function attains its minimum value, i.e. $A(r_{min})$.}
\label{figg3y}
\end{figure}

{We now discuss the configuration characterized by a null topological charge
in the presence of the same impurity already defined in the Eq. (\ref{mim}).
In the limit $d\rightarrow \infty $ and $c\rightarrow \infty $, with the
ratio $c/d=\gamma >0$ fixed (i.e.\ $c=\gamma {d}$}), we get that{%
\begin{equation}
\lim_{d\rightarrow \infty }\gamma {d}\exp \left( -dr^{2}\right) =\gamma \pi
\delta (r)\text{.}  \label{delta}
\end{equation}%
}

{In the Ginzburg-Landau model, at critical coupling, with a $\delta $%
-function impurity and }$\gamma =4${, the vortex with null} topological
charge behaves as a charge one vortex at critical coupling \cite{19}.
Surprisingly, this remains true also for axially symmetric configurations
away from critical coupling \cite{20}.

{The Figure \ref{figg3x} shows the }functions $\alpha (r)$ and $2+A(r)${\
for $c=4d$ and $d=1,~2,~4$ and $256$}. In general, as $d$ increases, the
shape of $\alpha (r)$ approaches that of a vortex with $m=1$, while the
minimum of the gauge field $A(r)$ tends to $-2$ and moves towards the origin
(see the Fig. \ref{figg3y}). This can be compared to the gauge field of a
vortex with $m=1$ shifted by $-2$. Hence, we have showed numerically that
the $\delta $-function impurities \textquotedblleft
behave\textquotedblright\ like vortices also in this more sophisticated
gauged-$CP(2)$ scenario. Moreover, while the profile function $\alpha (r)$
remains smooth, the gauge function $A(r)$ becomes singular in the limit $%
d\rightarrow \infty $ and develops a jump at the origin. {We have also
observed that, in general, when $c=4md$, the limit }$d\rightarrow \infty ${\
corresponds to a $\delta $-function of strength $4m\pi $ and the
corresponding solution approaches a topological profile (in the lack of the
impurity) with winding number $m$. Also in this case, the gauge profile
function $A(r)$ maintains the jump at $r=0$.}

The Figure \ref{figg3y} shows how the value of $\alpha (r=0)$ (dotted golden
line) goes to zero for large values of $d$, for $m=0$ and $c=4d$. The Figure
also shows how the value of $-A(r_{\text{min}})$ approaches $2$ whereas its
localization $r_{\text{min}}$ goes to zero (giving rise to the jump
mentioned previously) for large values of $d$. This fact justifies the plot
of $2+A(r)$ in the bottom of the Fig. \ref{figg3x}.


\section{The effective Chern-Simons-$CP(2)$ model \label{sec30}}

We {now} present the model {which} describes the interaction between the $%
CP(2)$ field and {the} Chern-Simons Abelian gauge one. {The model} {is}
defined by the Lagrange density%
\begin{equation}
\mathcal{L}=-\frac{\kappa }{4}\epsilon ^{\alpha \mu \nu }A_{\alpha }F_{\mu
\nu }+\left\vert \nabla _{\mu }\phi \right\vert ^{2}-V_{0}\left( \phi
_{3}\right) \text{,}  \label{ln0}
\end{equation}%
where the Chern-Simons term {controls} the gauge field's dynamics, {while}
the parameter $\kappa $ stands for the respective coupling constant. The
basic definitions, conventions and correlated discussions introduced in the
previous Sec. \ref{general0} remain the same.

For the configuration (\ref{beta1}), the effective model {then reads}%
\begin{eqnarray}
\mathcal{L} &=&-\frac{\kappa }{4}\epsilon ^{\alpha \mu \nu }A_{\alpha
}F_{\mu \nu }+\left\vert D_{\mu }\phi \right\vert ^{2}  \notag \\[0.2cm]
&&-V_{0}\left( \phi _{3}\right) -\lambda (h-\phi ^{\dag }\phi )\text{,}
\label{ln01}
\end{eqnarray}

We are interested in the BPS structure {which arises} from {the} model {above%
}. {In view of} the Gauss law%
\begin{equation}
\kappa B=-g^{2}A_{0}\left\vert \psi \right\vert ^{2}=-\frac{1}{2}%
g^{2}A_{0}\left( h-\phi _{3}^{2}\right) \text{,}  \label{CSL03}
\end{equation}%
the energy density {takes the form}%
\begin{equation}
\varepsilon =\frac{\kappa ^{2}B^{2}}{g^{2}\left( h-\phi _{3}^{2}\right) }%
+\left( D_{k}\phi \right) ^{\dag }D_{k}\phi +V_{0}\text{.}
\end{equation}

After some algebraic manipulations, the total {energy of the effective}
system becomes%
\begin{eqnarray}
\mathcal{E} &=&\int d^{2}\mathbf{x}\left\{ \frac{1}{2}\left\vert D_{j}\phi
\pm i\epsilon _{jk}h^{-1/2}\left( \phi \times D_{k}\phi \right) ^{\ast
}\right\vert ^{2}\right.  \notag \\[0.2cm]
&&\hspace{1cm}+\left( \frac{\kappa B}{g\left( h-\phi _{3}^{2}\right) ^{1/2}}%
\mp \sqrt{V_{0}}\right) ^{2}\pm 2\pi h\bar{q}_{0}  \notag \\[0.2cm]
&&\hspace{1cm}\left. \pm B\left( \frac{2\kappa \sqrt{V_{0}}}{g\left( h-\phi
_{3}^{2}\right) ^{1/2}}-h^{1/2}g\phi _{3}\right) \right\} \text{,}\quad \quad
\end{eqnarray}%
where $\bar{q}_{0}$ is the topological charge density defined in {the
previous} Eq. (\ref{TPcharge}). {Again in this case,} {whether we choose}\
the factor {which multiplies} the magnetic field {as being equal to zero},
we determine the BPS potential of the model (\ref{ln01}), {i.e.}%
\begin{equation}
V_{0}=\frac{hg^{2}}{4\kappa ^{2}}\phi _{3}^{2}(h-\phi _{3}^{2})\text{.}
\end{equation}%
{via which} we {complete} the implementation of the BPS formalism for the
model (\ref{ln01}) {by writing the total energy as}%
\begin{eqnarray}
\mathcal{E} &=&\mathcal{E}_{bps}+\int d^{2}\mathbf{x}\left( \frac{\kappa B}{%
g\left( h-\phi _{3}^{2}\right) ^{1/2}}\mp \sqrt{V_{0}}\right) ^{2}  \notag \\%
[0.2cm]
&&+\frac{1}{2}\int d^{2}\mathbf{x}\left\vert D_{j}\phi \pm ih^{-1/2}\epsilon
_{jk}\left( \phi \times D_{k}\phi \right) ^{\ast }\right\vert ^{2}\!\!\text{,}\quad
\end{eqnarray}%
where $\mathcal{E}_{bps}$ is the same {one already} defined in {the} Eq. (%
\ref{ebpssf}).

We see that total energy becomes equal to $\mathcal{E}_{bps}$ {when} the
quadratic terms {within} the integrals {are assumed to be zero}, {from which
one gets} the BPS or self-dual equations of the system, i.e.%
\begin{equation}
B=\pm \frac{h^{1/2}g^{2}}{2\kappa ^{2}}\phi _{3}(h-\phi _{3}^{2})\text{,}
\label{csBPS01}
\end{equation}%
\begin{equation}
D_{j}\phi =\mp ih^{-1/2}\epsilon _{jk}\left( \phi \times D_{k}\phi \right)
^{\ast }\text{,}  \label{csBPS02}
\end{equation}%
{which mimic} {the ones inherent to} the Chern-Simons-$O(3)$ sigma model. {%
Furthermore,} according to {the refs.} \cite{witten, spector}, the BPS
system {above} is related to an extended supersymmetric {version} of the
model (\ref{ln0}).

In {the} Ref. \cite{cscp2}, {the authors} studied the rotationally symmetric
solutions of the BPS system (\ref{csBPS01}) and (\ref{csBPS02}), {for} the
case $\beta =\beta _{1}$. In the next Section, we {investigate} the effects
of a magnetic impurity on the BPS solitons supported by the model (\ref{ln01}%
).


\section{Chern-Simons-$CP(2)$ vortex-like solitons in the presence of a
magnetic impurity\label{sec3}}

We now consider a second enlarged model which describes the interaction
between the $CP(2)$-field and a Chern-Simons Abelian gauge one (i.e. a
Chern-Simons-$CP(2)$ model). The resulting model is defined by the Lagrange
density
\begin{equation}
{\mathcal{L}}=-\frac{\kappa }{4}\epsilon ^{\alpha \mu \nu }A_{\alpha }F_{\mu
\nu }+\left| D_{\mu }\phi \right| ^2-V\left( \phi _{3},\Delta \right)
+\Delta B.  \label{ln}
\end{equation}

Also in this Section, our study focuses on those time-independent
configurations with radial symmetry. With such a purpose in mind, we again
use the map defined by Eqs. (\ref{2m}) and (\ref{3m})for the profile
functions $\alpha (r)$ and $A(r)$ that still obey the boundary conditions (%
\ref{31m}) and (\ref{bc2m}). Besides, the scalar potential $A_{0}$ is also
supposed to depend on the radial coordinate $r$ only,
\begin{equation}
A_{0}=A_{0}(r)\text{,}
\end{equation}%
while the expressions for the magnetic and electric fields are%
\begin{equation}
B(r)=-\frac{1}{gr}\frac{dA}{dr}\text{ \ and \ }E(r)=-\frac{dA_{0}}{dr}\text{,%
}  \label{be}
\end{equation}%
respectively.

Here, as in the previous model, the term $\Delta B$ does not change the
Gauss law (\ref{CSL03}) which comes from Lagrange density (\ref{ln0}) when
considered in the absence of the impurity. So, one gets the Gauss law as
\begin{equation}
\kappa B=-\frac{g^{2}h}{2}A^{0}\sin ^{2}\alpha \text{,}  \label{gl1}
\end{equation}%
from which we get that the {new model possesses configurations which} carry
both magnetic flux and electric charge simultaneously, a well-known effect
caused by the presence of the Chern-Simons term itself (see Ref. \cite{cscp2}
and the discussion therein).

In what follows, we again focus our attention on those first-order solutions
which minimize the total energy of the model. With such a purpose in mind,
we implement the Bogomol'nyi prescription, the starting-point being the
radially symmetric expression for the energy density, which we write in a
more convenient form as
\begin{eqnarray}
\varepsilon &=&\frac{\kappa ^{2}B^{2}}{g^{2}h\sin ^{2}\alpha }+V(\alpha
,\Delta )-\Delta B  \notag \\[0.2cm]
&&+h\left[ \left( \frac{d\alpha }{dr}\right) ^{2}+\frac{\left( 2m-A\right)
^{2}}{4r^{2}}\sin ^{2}\alpha \right] \text{,}  \label{oe}
\end{eqnarray}%
where we have used the Gauss law (\ref{gl1}) to express the scalar potential
$A_{0}(r)$ as a function of the magnetic field $B(r)$.

{Now, from the Eq. (\ref{oe}), we initiate the implementation of the BPS
technique providing, after some algebra, the following expression for the
total energy:
\begin{eqnarray}
\frac{\mathcal{E}}{2\pi } &=&\int_{0}^{\infty }\left[ h\left( \frac{d\alpha
}{dr}\pm \frac{\left( 2m-A\right) }{2r}\sin \alpha \right) ^{2}\right.
\notag \\
&&\hspace{0cm}+\left( \frac{\kappa B}{g\sqrt{h}\sin \alpha }\mp \sqrt{V}%
\right) ^{2}\pm 2\pi h\bar{q}_{0}  \notag \\
&&\hspace{0cm}\left. \pm B\left( \frac{2\kappa \sqrt{V}}{gh^{1/2}\sin \alpha
}-hg\cos \alpha \mp \Delta \right) \right] rdr\text{,}\quad   \label{eeb1}
\end{eqnarray}%
where the quantity $\bar{q}_{0}$ is the topological charge density of the
model, being the same given in Eq. (\ref{qq0}).}

To complete the minimization of the total energy according to the
Bogomol'nyi prescription, we set to zero the expression that multiplies the
magnetic field in the third row of Eq. (\ref{eeb1}) it allow us to determine
the BPS potential $V(\alpha ,\Delta )$ as
\begin{equation}
V(\alpha ,\Delta )=\frac{g^{4}h^{3}}{4\kappa ^{2}}\left( \cos \alpha \pm
\frac{\Delta }{gh}\right) ^{2}\sin ^{2}\alpha \text{,}  \label{rsp}
\end{equation}%
where both the potential and the function $\Delta $ go to zero when $%
r\rightarrow \infty $.

This way, the total energy becomes written in the form%
\begin{eqnarray}
\frac{\mathcal{E}}{2\pi } &=&\frac{\mathcal{E}_{bps}}{2\pi }%
+\int_{0}^{\infty }\left[ \left( \frac{\kappa B}{g\sqrt{h}\sin \alpha }\mp
\sqrt{V}\right) ^{2}\right] rdr  \notag \\[0.2cm]
&&\hspace{0cm}+h\int_{0}^{\infty }\left[ \frac{d\alpha }{dr}\pm \frac{\left(
2m-A\right) }{2r}\sin \alpha \right] ^{2}rdr\text{,}  \label{eeb2}
\end{eqnarray}%
where we have introduced the energy $\mathcal{E}_{bps}$ defined in Eq. (\ref%
{ebpssf}) and whose value is the same from Eq. (\ref{8m}). Furthermore, from
the Eq. (\ref{eeb2}), we write the inequality
\begin{equation}
\mathcal{E}\geq \mathcal{E}_{bps}\text{,}  \label{bbond}
\end{equation}%
{from which we clearly see that $\mathcal{E}_{bps}$ stands for the
Bogomol'nyi bound which can be calculated in the very same way }as before
(i.e. via the usage of the boundary conditions (\ref{31m}) and (\ref{bc2m}%
)). The inequality (\ref{bbond}) reveals that the Bogomol'nyi bound is
saturated when the fields which appear in the Eq. (\ref{eeb2}) satisfy the
BPS equations:%
\begin{equation}
B=\pm \frac{g^{3}h^{2}}{2\kappa ^{2}}\left( \cos \alpha \pm \frac{\Delta }{gh%
}\right) \sin ^{2}\alpha \text{,}  \label{xxb1}
\end{equation}%
\begin{equation}
\frac{d\alpha }{dr}=\mp \frac{\left( 2m-A\right) }{2r}\sin \alpha \text{,}
\label{xxb2}
\end{equation}%
whose solutions describe time-independent configurations with total energy
given by $\mathcal{E}=\mathcal{E}_{bps}=4\pi h\left\vert m\right\vert $,
which is equal to the energy inherent to the BPS structures obtained in the
previous Maxwell-$CP(2)$ case (see the Sec. \ref{sec2} and the discussion
therein).

Furthermore, the BPS energy density obtained from Eq. (\ref{oe}) is
\begin{equation}
\varepsilon _{bps}=\left( \sqrt{2V}\mp \frac{\Delta }{2}\right) ^{2}-\frac{%
\Delta ^{2}}{4}+2h\left( \frac{d\alpha }{dr}\right) ^{2}\text{,}
\label{EEbps1}
\end{equation}%
with the BPS potential $V$ given by the Eq. (\ref{rsp}). The last can be
rewritten in terms of $\phi _{3}$ as%
\begin{equation}
V\left( \phi _{3},\Delta \right) =\frac{g^{4}h}{4\kappa ^{2}}\left( \phi
_{3}\pm \frac{\Delta }{g\sqrt{h}}\right) ^{2}\left( h-\phi _{3}^{2}\right)
\text{,}
\end{equation}
which allows the spontaneous breaking of the $SU(3)$ symmetry inherent to
the original Chern-Simons-$CP(2)$ model, as expected.

We investigate below the first-order equations (\ref{xxb1}) and (\ref{xxb2})
numerically. In the sequence, we plot the resulting BPS profiles and comment
on their main properties engendered by the presence of a localized impurity.

\subsection{The Chern-Simons-$CP(2)$ vortex-like solitons: numerical results}

In the sequence, we choose the localized magnetic impurity as in the
previous Eq. (\ref{mim}), i.e.%
\begin{equation}
\Delta (r)=c e^{-dr^2} \text{,}  \label{DDt}
\end{equation}
via which we rewrite the potential (\ref{rsp}) in the form
\begin{equation}
V(\alpha ,\Delta )=\frac{g^{4}h^{3}}{4\kappa ^{2}}\left( \cos \alpha \pm %
\displaystyle\frac{c}{gh}e^{-dr^{2}} \right) ^{2}\sin ^{2}\alpha \text{.}
\end{equation}

{In this sense, the BPS equations (\ref{xxb1}) and (\ref{xxb2}) become}%
\begin{equation}
\frac{1}{r}\frac{dA}{dr}=\mp \frac{g^{4}h^{2}}{2\kappa ^{2}}\left( \cos
\alpha \pm \frac{c}{hg}e^{-dr^{2}}\right) \sin ^{2}\alpha \text{,}
\label{x2}
\end{equation}%
\begin{equation}
\frac{d\alpha }{dr}=\mp \frac{\left( 2m-A\right) }{2r}\sin \alpha \text{,}
\label{x1}
\end{equation}%
{whose solutions must satisfy the boundary conditions (\ref{31m}) and (\ref%
{bc2m})}. As in the previous Sec. II, {we only consider the lower signs in
the BPS equations in order to describe the first-order solutions for $m>0$.}

\begin{figure}[tbp]
\includegraphics[width=8.4cm]{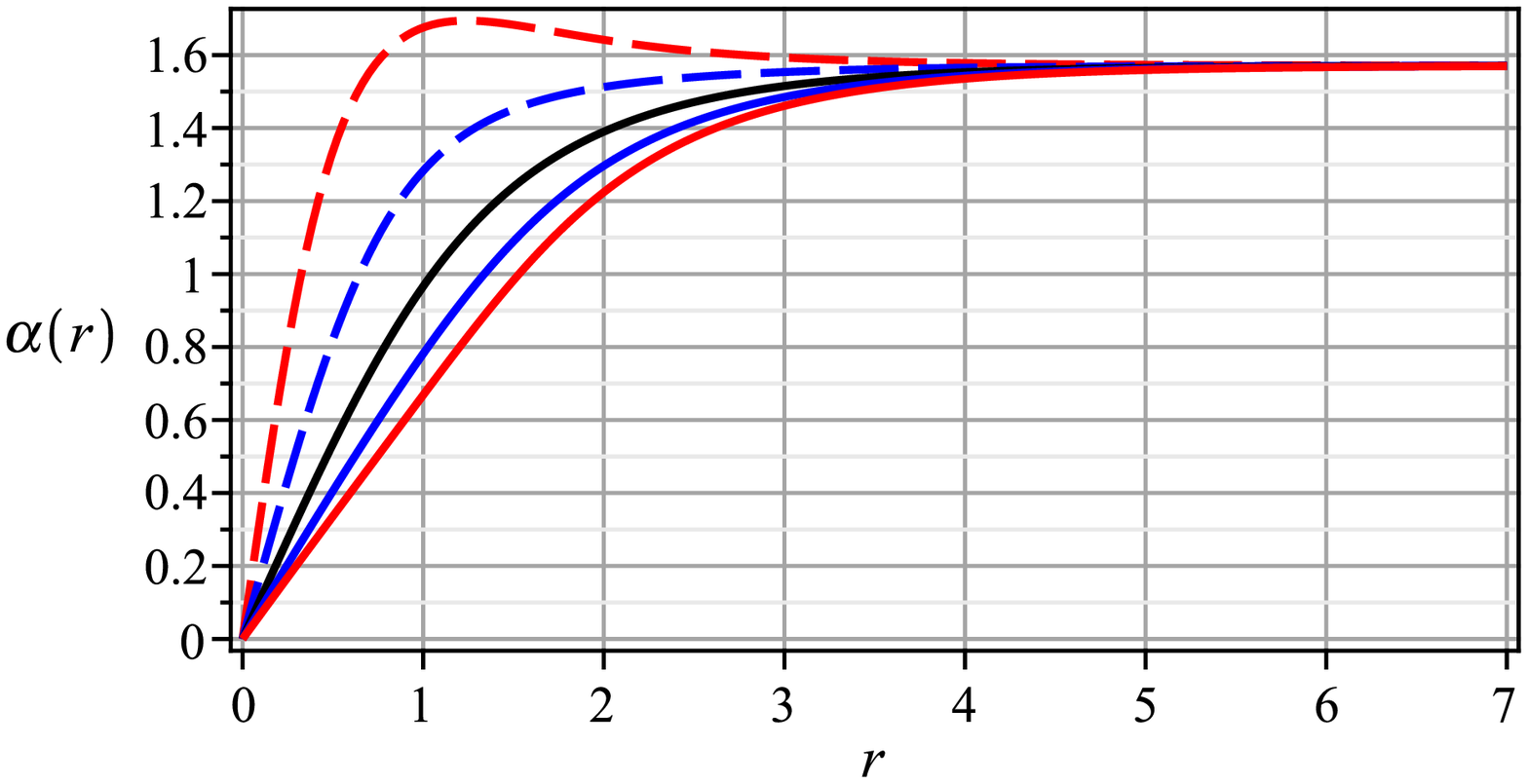} %
\includegraphics[width=8.4cm]{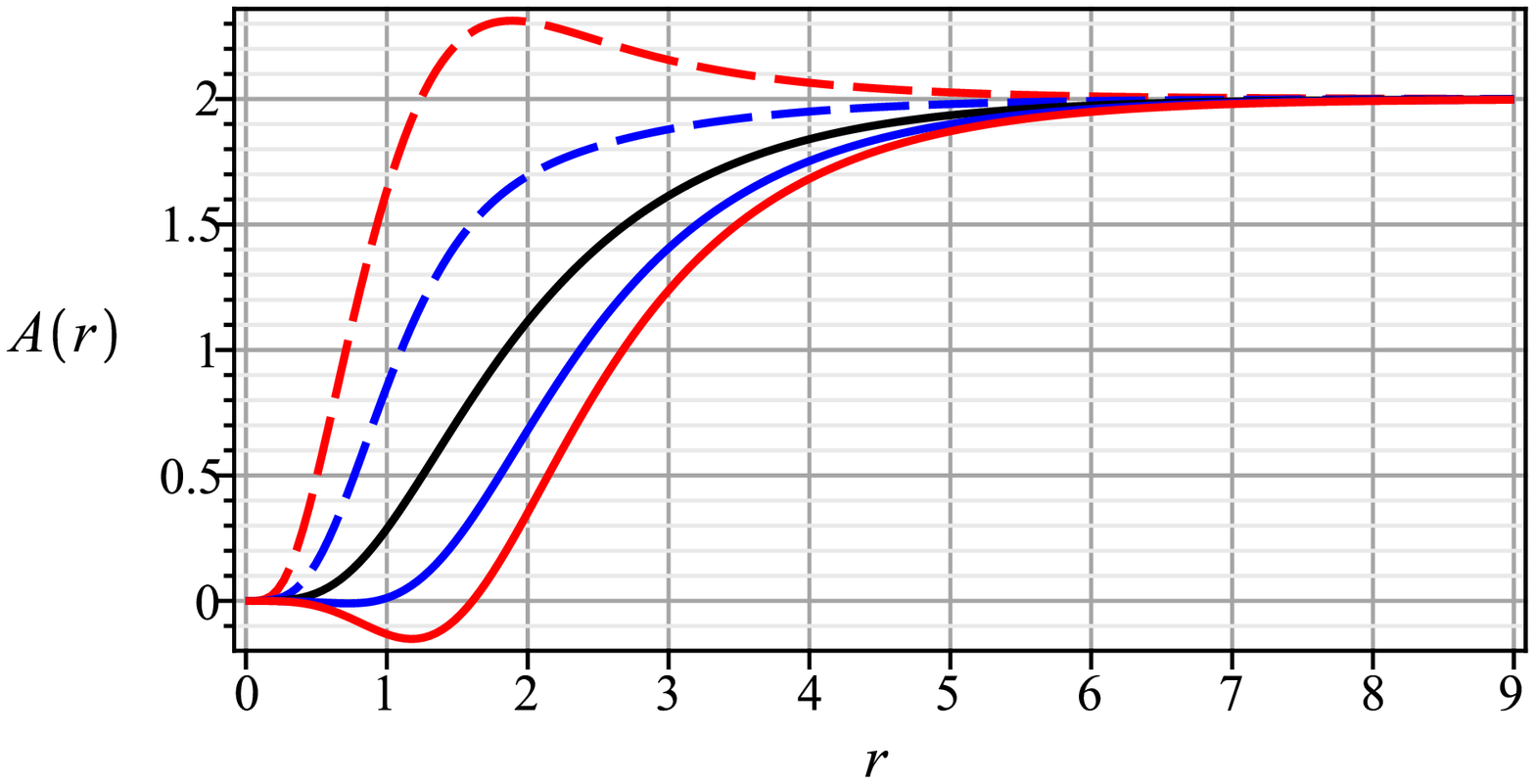}
\caption{Numerical solutions to $\protect\alpha \left( r\right) $ (top) and $%
A(r)$ (bottom) coming from (\protect\ref{x2}) and (\protect\ref{x1}) in the
presence of (\protect\ref{31m}) and (\protect\ref{bc2m}). {The results hold
for }$\protect\kappa =h=m=1${, $g=\protect\sqrt{2}$, $d=1$ and }$c=-4$
(dashed red line), $c=-2$ (dashed blue line), $c=0$ (solution without
magnetic impurity, solid black line), $c=+2$ (solid blue line) and $c=+4$
(solid red line).}
\label{figg4}
\end{figure}

\begin{figure}[tbp]
\includegraphics[width=8.4cm]{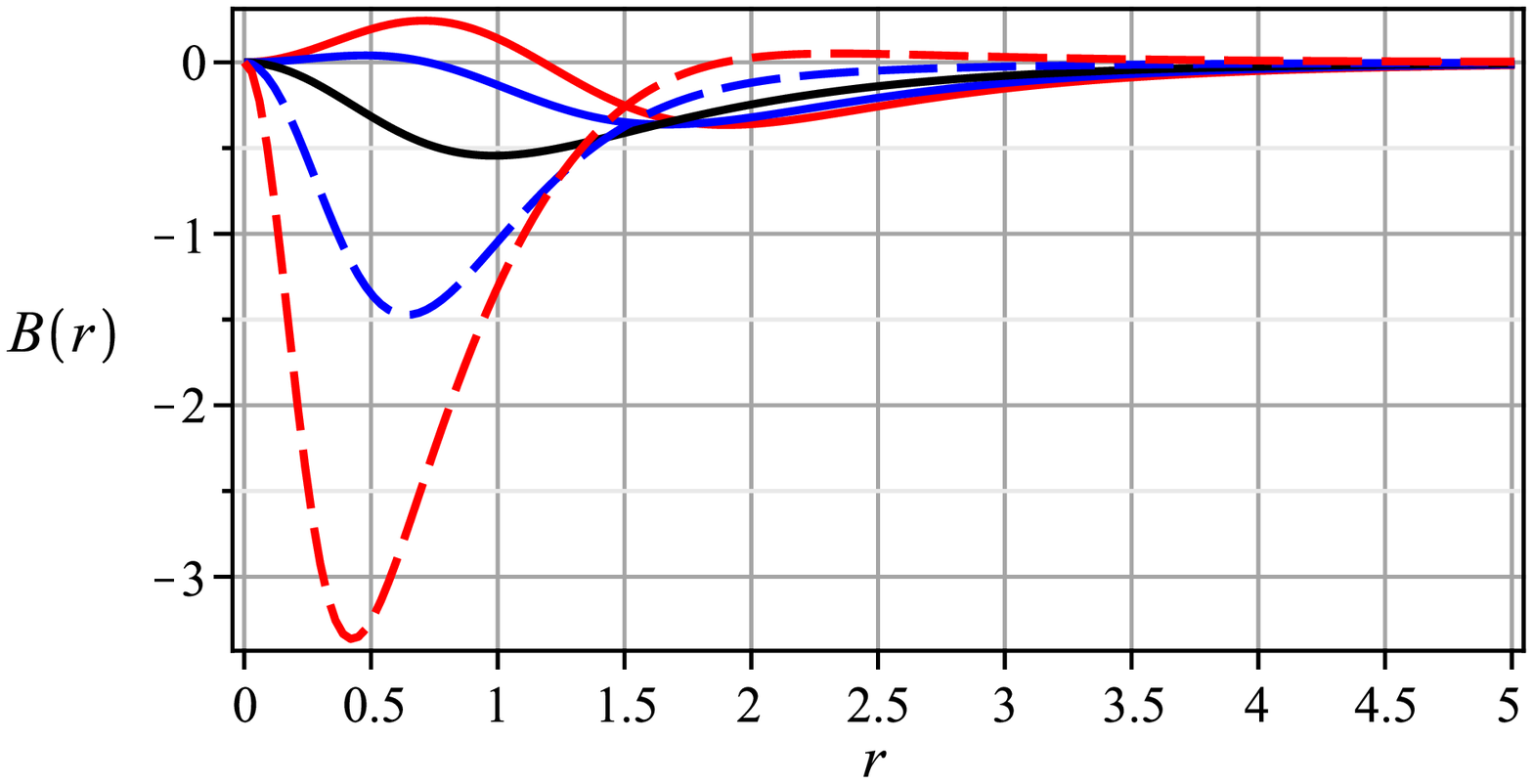} %
\includegraphics[width=8.4cm]{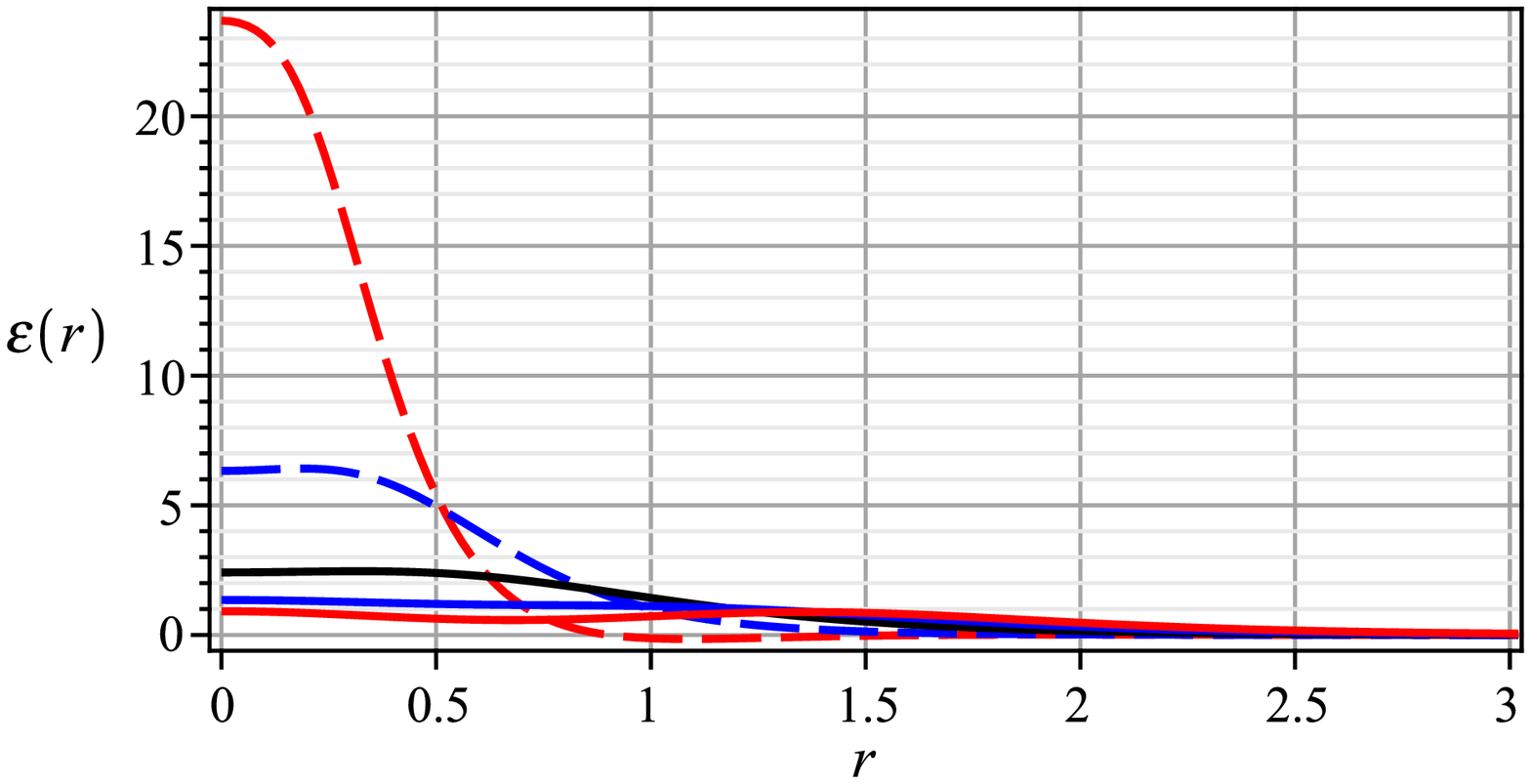}
\caption{Numerical solutions to the magnetic field $B\left( r\right) $ (top)
and the energy density $\protect\varepsilon _{bps}\left( r\right) $ (bottom)
of the first-order Chern-Simons-$CP(2)$ configurations. Conventions as in
the Fig. \protect\ref{figg4}.}
\label{figg5}
\end{figure}

\begin{figure}[tbp]
\includegraphics[width=8.4cm]{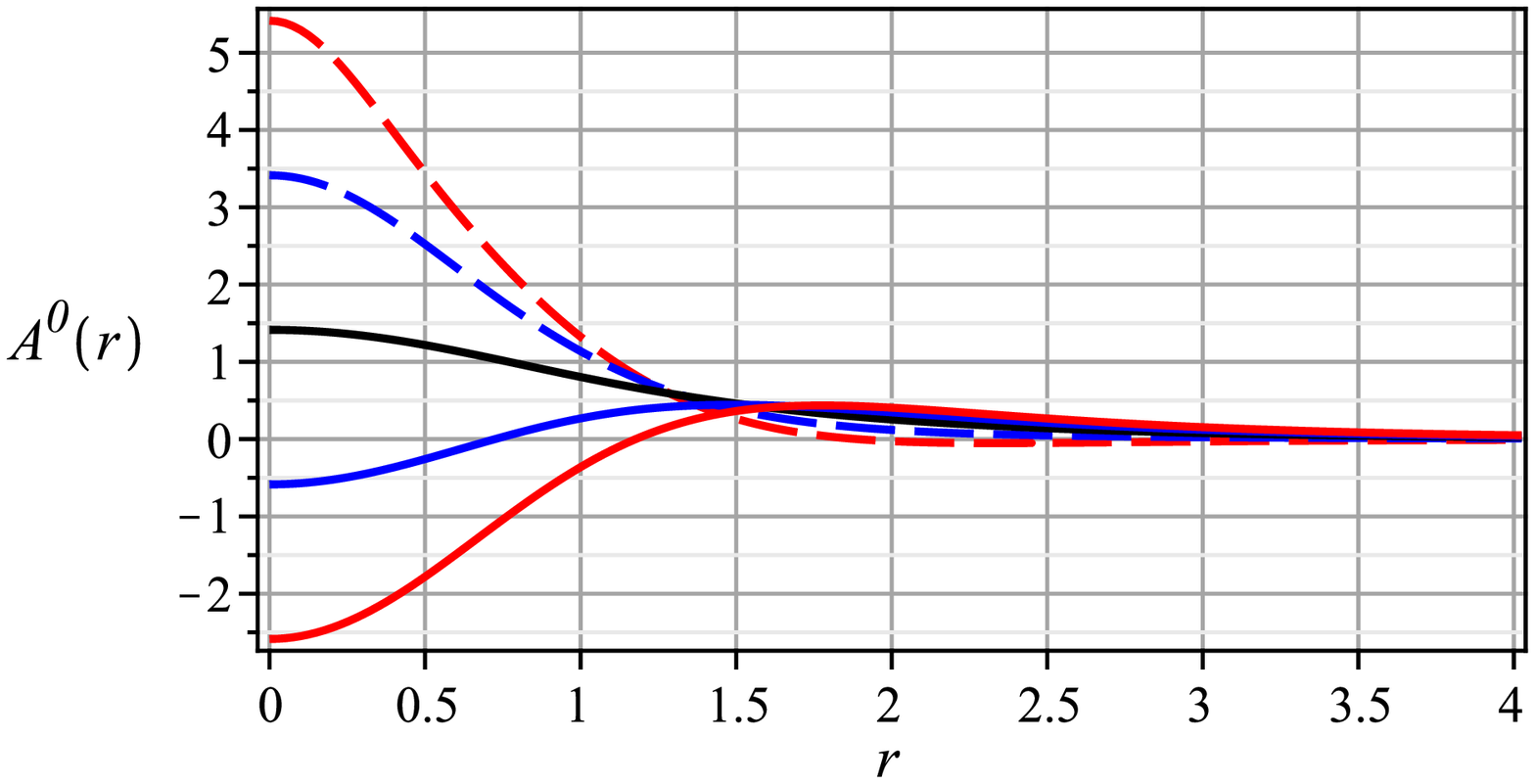} %
\includegraphics[width=8.4cm]{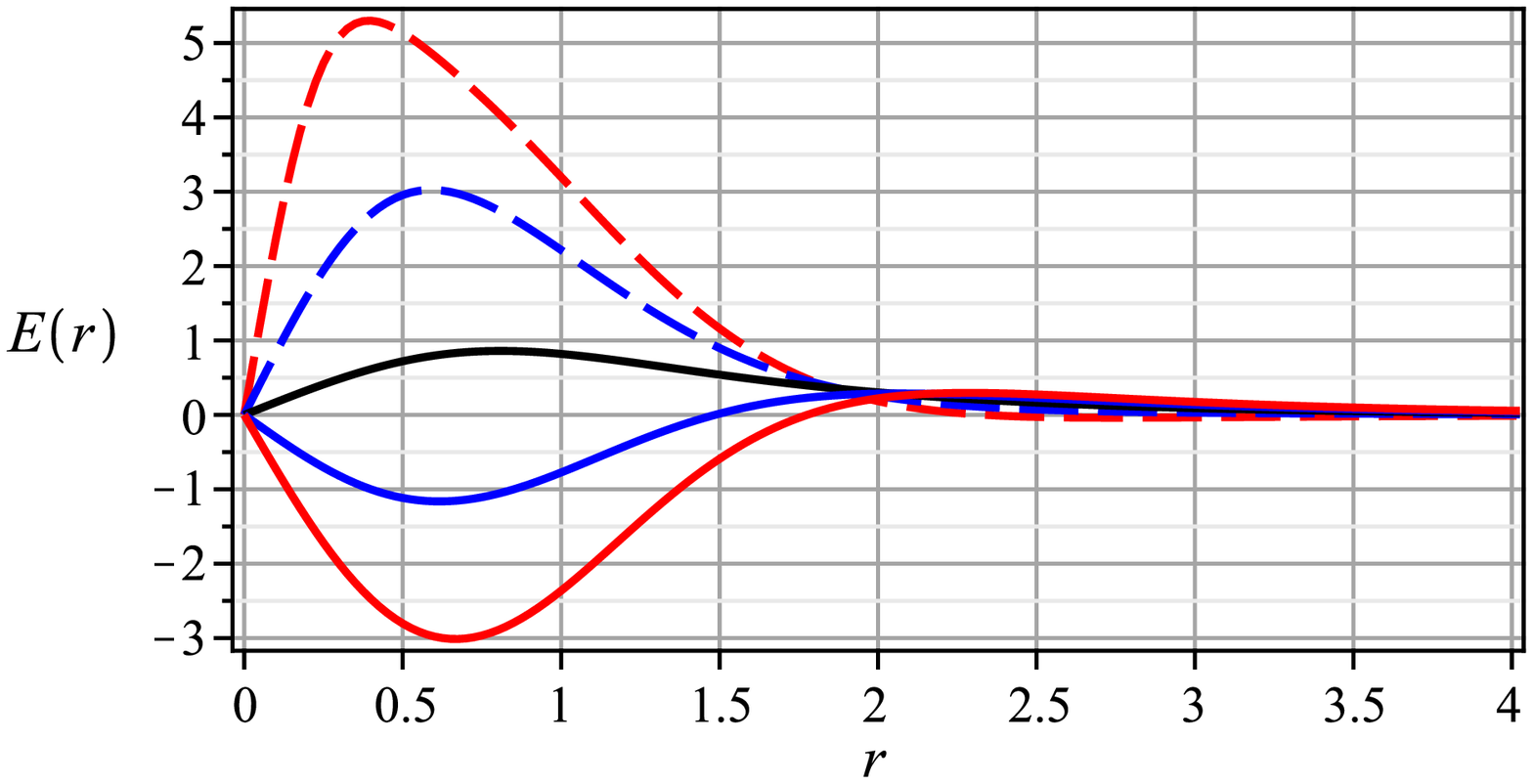}
\caption{Numerical solutions to the scalar potential {$A^{0}(r)$} (top) and
the electric field $E(r)$ (bottom) of the first-order Chern-Simons-$CP(2) $\
configurations. Conventions as in the Fig. \protect\ref{figg4}.}
\label{figg6}
\end{figure}

We again implement a finite-difference algorithm in order to {solve }the
first-order equations (\ref{x2}) and (\ref{x1}) {numerically}. In this
sense, we choose $\kappa =h=1$, $g=\sqrt{2}$, $m=1$, $d=1$ (the impurity's
\textquotedblleft width"), from which we study the {BPS configurations} for
the same values of $c$ (the impurity's \textquotedblleft height") already
considered in the previous Sec. \ref{sec2}, i.e. $c=-4$ (dashed red line), $%
c=-2$ (dashed blue line), $c=0$ ({usual solution, no impurities,} solid
black line), $c=+2$ (solid blue line) and $c=+4$ (solid red line). {We
depict the numerical solutions for the relevant fields in the figures} \ref%
{figg4}, \ref{figg5} and \ref{figg6}. Here, it is important to say that the
solution for $c=-5$\ are not shown because the effects caused by the
impurity can be seen clearly through the profile for $c=-4$.

{The Figure \ref{figg4} brings the solutions to the profile functions $%
\alpha (r)$ and $A(r)$, from which one notes }that the same effects are
again present when the values of $|c|$ increase, {{i.e. the profiles lose
their monotonicity because of the presence of the magnetic impurity.} In
particular, due to the loss of monotonicity, the $\alpha (r)$-profiles can
assume values {that are eventually bigger }than $\pi /2$.}

{The Figure \ref{figg5} depicts} the magnetic field $B(r)$ and the energy
density $\varepsilon _{bps}$. For increasing values of $|c|$, we again
identify an inversion on the sign (i.e. a flip) of the magnetic field {as
already observed in the previous model, such an effect} being caused by the
magnetic impurity which gives rise to a gauge field with a nonmonotonic
shape. Moreover, {despite the effects caused by the impurity, the energy
density remains} localized and well-behaved along the radial coordinate, as
expected.

{Finally, {we plot} the profiles for the scalar potential $A^{0}(r)$ and the
electric field $E(r)=-dA_{0}/dr$ {in the Figure \ref{figg6}. }The Gauss law (%
\ref{gl1}) defines a linear dependence between the scalar potential and the
magnetic field, which means that the flipping of $B$ {(see the Fig. 5)}
leads to an inversion on the sign of $A^{0}$, and vice-versa. Hence, {the
scalar potential's nonmonotonic behavior also produces the sign inversion of
the electric field itself.}}

\subsubsection{Behavior of the solutions near the origin}

We now investigate the way the profile fields $\alpha \left( r\right) $ and $%
A(r)$ approach the values (\ref{31m}) and (\ref{bc2m}). Without loss of
generality, we consider only those configurations with positive values of
the winding number $m$. Thus, the behaviors of the profile functions near
the origin are
\begin{eqnarray}
&\displaystyle \alpha (r)\approx \mathcal{C}_{0}r^{m}\text{,} &  \label{a1}
\\[0.3cm]
&\displaystyle A(r)\approx \frac{\mathcal{C}_{0}^{2}g^{3}h\left( hg-c\right)
}{4\kappa^{2}\left( m+1\right) }r^{2(m+1)}\text{,} &  \label{a2}
\end{eqnarray}
where $\mathcal{C}_{0}$ stands for a positive real constant.

{We write below\ the behaviors near the origin for the magnetic field, the
BPS energy density, the scalar potential and the electric field. For such a
purpose, we consider }$\kappa =h=1${, $g=\sqrt{2}$, $m=1$, and $d=1$, i.e.
the same values used to obtain the previous numerical solutions. Then, we
get the following behavior for the magnetic sector,%
\begin{equation}
B(r)\approx \mathcal{C}_{0}^{2}(c-\sqrt{2}){r}^{2}\text{,}
\end{equation}%
and for the BPS energy density, we obtain%
\begin{equation}
\varepsilon _{bps}(r)\approx 2\mathcal{C}_{0}^{2}-\frac{\mathcal{C}_{1}}{3}{r%
}^{2}\text{,}
\end{equation}%
where $\mathcal{C}_{1}=\mathcal{C}_{0}^{2}(3\sqrt{2}\,c+2\mathcal{C}%
_{0}^{2}-6)$.}

As a result, {the expressions above offer an explanation in terms of the
values of }${c}${\ about the behavior of the corresponding sectors near the
origin as appear in the Fig. \ref{figg5}.}

Further, the behavior of both the scalar potential and electric field
becomes
\begin{eqnarray}
&\displaystyle {A}_{0}(r)\approx \sqrt{2}-c+\left( c-\frac{\sqrt{2}}{2}%
\mathcal{C}_{0}^{2}\right) {r}^{2}\text{,} & \\[0.2cm]
&\displaystyle {E}(r)\approx (\sqrt{2}\mathcal{C}_{0}^{2}-2c)r-\frac{\sqrt{2}
\mathcal{C}_{0}^{4}-6c}{3}{r}^{3}\text{,}&
\end{eqnarray}
respectively. These approximate solutions also explain the behaviors near to
$r=0$ depicted in the Fig. \ref{figg6}.

\begin{figure}[t]
\includegraphics[width=8.4cm]{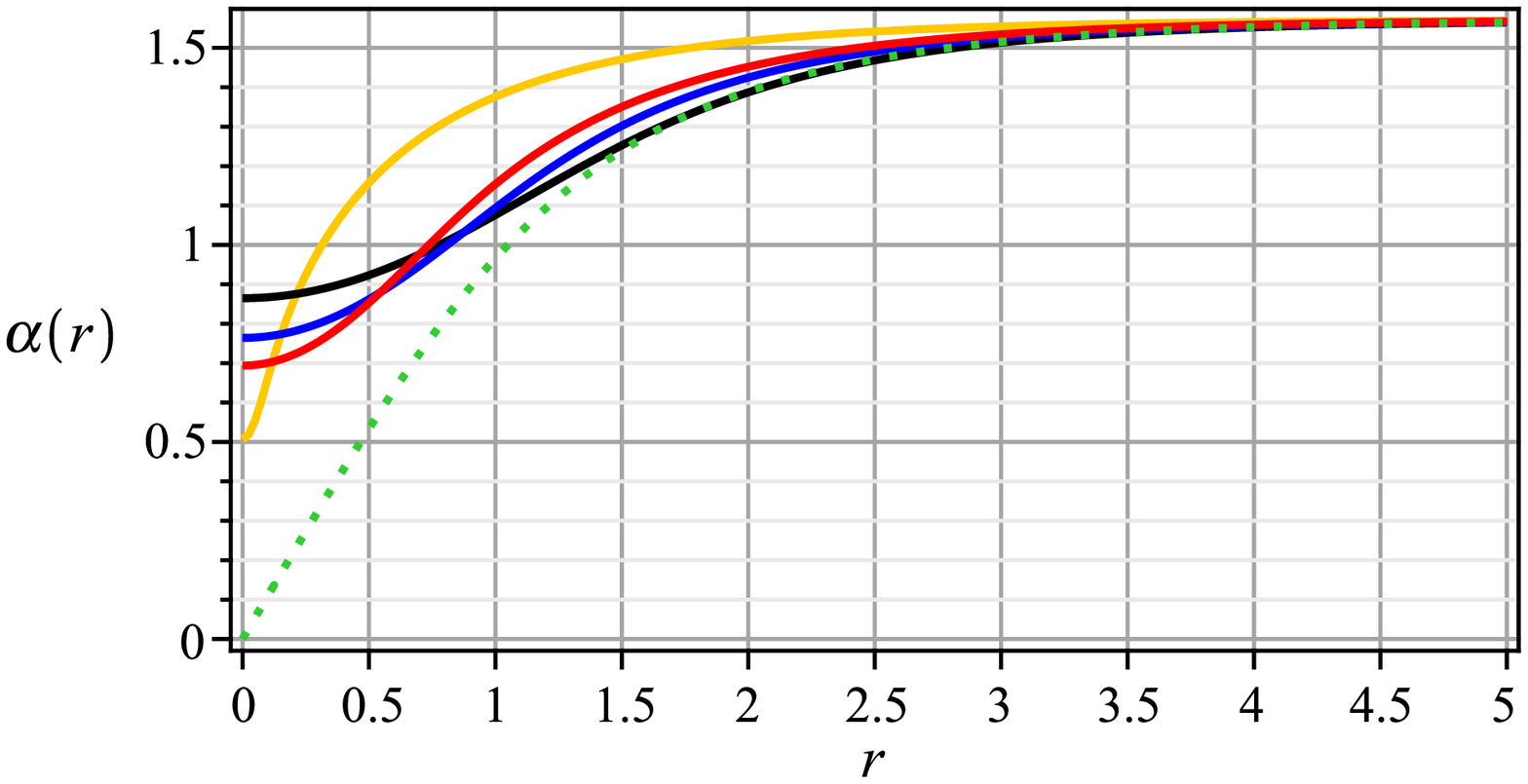} %
\includegraphics[width=8.4cm]{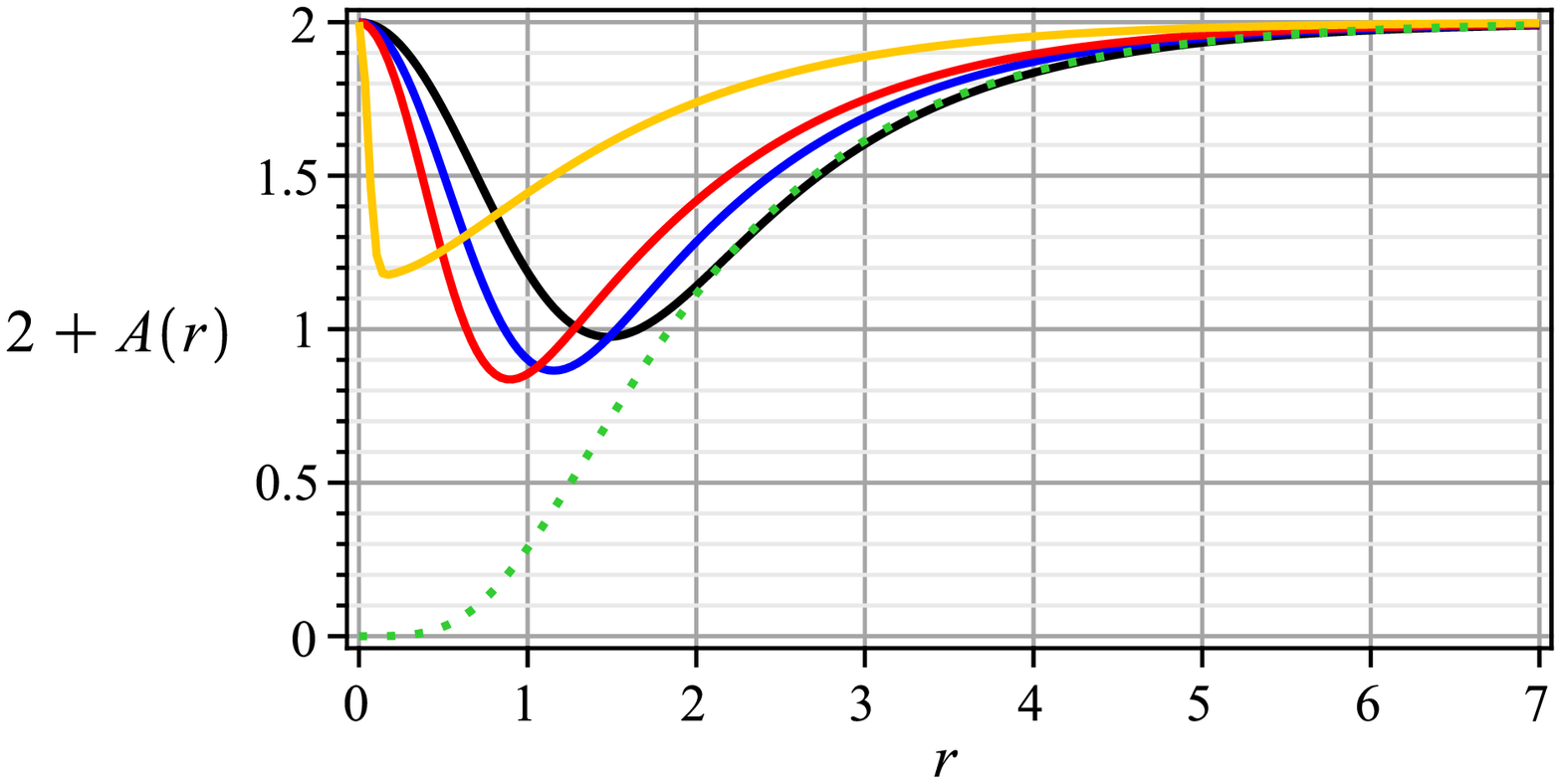}
\caption{Numerical solutions to $\protect\alpha (r)$ (top) and $2+A(r)$
(bottom) for $m=0$ (i.e. the vacuum configuration). The impurity is still
given by the Eq. \eqref{DDt}. Here, we have depicted the solutions for $d=1$
($c=4$, black line), $d=2$ ($c=8$, blue line), $d=4$ ($c=16$, red line) and $%
d=256$ ($c=1024$, orange line). Again, the dotted green line stands for the
topological profile with $m=1$\ in the absence of impurities.}
\label{figg7}
\end{figure}

\begin{figure}[t]
\includegraphics[width=8.4cm]{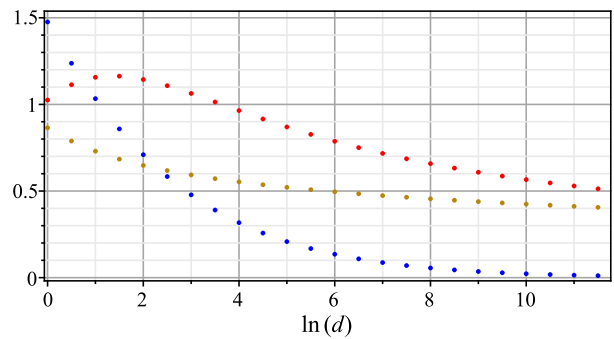}
\caption{The Chern-Simons-$CP(2)$ case:\ this Figure shows the behavior of $%
-A(r_{min})$ (dotted red line), $r_{min}$ (dotted blue line) and $\protect%
\alpha (r=0)$ (dotted golden line) \textit{versus} ln$(d)$. As in the
previous case, $r_{min}$ is the value of $r$ for which the gauge profile
function attains its minimum value.}
\label{figg7x}
\end{figure}

\subsubsection{Behavior of the solutions in the asymptotic limit}

On the other hand, in the asymptotic limit $r\rightarrow \infty $, the
fields behave as%
\begin{eqnarray}
&\displaystyle\alpha (r)\approx \frac{\pi }{2}-\mathcal{C}_{\infty }\frac{%
\exp \left( -\mathcal{M}r\right) }{\sqrt{r}}\text{,}& \\[0.3cm]
&\displaystyle A(r)\approx 2m-2\mathcal{M}\mathcal{C}_{\infty }\sqrt{r}\exp
\left( -\mathcal{M}r\right) \text{,}&
\end{eqnarray}%
{for all the values of $c$ and $d$ in (\ref{DDt})}, where $C_{\infty }$
stands for a positive real constant and
\begin{equation}
\mathcal{M}=\frac{g^{2}h}{2\kappa }
\end{equation}%
represents the mass of both the bosons in the Bogomol'nyi limit. As in the
Maxwell-$CP(2)$ case, it is possible to conclude also in the present
Chern-Simons scenario that a localized impurity does not change the way the
profile functions approach their asymptotic values.

\subsection{The topologically trivial solution: the $m=0$ case}

We finally end this Section by considering the nontopological ($m=0$)
configuration engendered by the $\delta$-function impurity given in the Eq. (%
\ref{delta}), see the Sec. \ref{sec2b} and the discussion therein.

The numerical results for the profile functions $\alpha (r)$ and $2+A(r)$
are shown in the Fig. \ref{figg7}, again for $\gamma =4$ and $d=1$,$~2$,$~4$
and $256$. We observe that in the present case the $m=0$-solution does not
mimic the behavior previously found in the Maxwell-$CP(2)$ scenario, i.e.
the profile for $\alpha(r)$ does not approach that of a $m=1$ topological
vortex as $d$ increases. At the same time, the minimum of $A(r)$ does not
saturate to the value $-2$ whereas it moves towards the origin, see the Fig. %
\ref{figg7x}. As a consequence, in the present Chern-Simon-$CP(2)$ case, the
topologically trivial configuration does not behave as a $m=1$-vortex, which
arises in the absence of the impurity.


\section{Final comments and perspectives\label{sec4}}

{We have performed the construction of BPS vortices in the context of two
different gauged-$CP(2)$ scenarios that {were enlarged} via an additional
term {which represents} a magnetic impurity. With such an aim {in mind}, we
have chosen a specific $CP(2)$-configuration (\ref{beta1}) {as being coupled
to }both the Maxwell's as the Chern-Simons fields, separately. {Here,} it is
worthwhile to {highlight that} such a configuration {presents} a $CP(2)$
topological charge {which equals zero not only in the simplest (free)
scenario, but also} when the $CP(2)$ {\ field is} coupled to an Abelian
gauge {one}. However, the BPS formalism shows that the effective models for (%
\ref{beta1}) possess a self-dual structure {which} looks like {that} of the
gauged sigma models. {Moreover,} the full implementation of the BPS
technique allows {us to fix }the self-dual potentials in both the Maxwell
and the Chern-Simons cases, {see} the {eqs.} (\ref{x14a}) and (\ref{rsp}),
respectively. {In this sense,} we {have verified} that the magnetic impurity
contributes {explicitly} to the self-dual potentials and appears in both
models' BPS equations. The interesting point is that the impurity does not
change the Bogomol'nyi bounds saturated by the BPS configurations, i.e. the
corresponding self-dual energies remain quantized according to the winding
number $m$, as expected for topological structures.}

{In order to study the {effects caused by }the magnetic impurity on the
solutions of the BPS systems, we have particularized our analysis by
choosing a Gaussian (localized) impurity controlled by two real parameters, $%
c$ and $d$ (which control the \textquotedblleft height" and the
\textquotedblleft width" {of the impurity}, respectively). For a fixed value
of $d$ and different values of $c$, the numerical analysis {has demonstrated
}how the parameter $c$ induces {not only} the loss of monotonicity of the {%
profile functions} $\alpha (r)$ and $A(r)$, {but also} the flipping of both
the magnetic and electric fields. The analysis of the behavior near the
origin {has explained} both peculiarities. In addition, we have verified
that the impurity does not change the manner these fields behave in the
asymptotic region ({i.e.} $r\rightarrow \infty $).}

{Based on the results which we have introduced in this work, an interesting
issue to be studied in the future is the effect eventually caused by a
localized impurity on the shape of nontopological BPS Chern-Simons-$CP(2)$
vortices. Another point which claims for a future analysis includes the
study of the interaction between a moving gauged-$CP(2)$ vortex and a static
magnetic impurity. The results of these topics, currently under
investigation, will be reported elsewhere.}

\begin{acknowledgments}
This work was financed in part by the Coordena\c{c}\~{a}o de Aperfei\c{c}%
oamento de Pessoal de N\'{\i}vel Superior - Brasil (CAPES) - Finance Code
001, the Conselho Nacional de Pesquisa e Desenvolvimento Cient\'{\i}fico e
Tecnol\'{o}gico - CNPq and the Funda\c{c}\~{a}o de Amparo \`{a} Pesquisa e
ao Desenvolvimento Cient\'{\i}fico e Tecnol\'{o}gico do Maranh\~{a}o -
FAPEMA (Brazilian agencies). In particular, VA thanks the full support from
CAPES (via a PhD scholarship). RC acknowledges the support from the grants
CNPq/306724/2019-7, CNPq/423862/2018-9, FAPEMA/Universal-01131/17 and
FAPEMA/Universal-00812/19. EH thanks the support from the grants
CNPq/307545/2016-4, CNPq/309604/2020-6 and FAPEMA/COOPI/07838/17. SK would
like to thank Jack McKenna and Abera Muhamed for interesting discussions. EH
also acknowledges the School of Mathematics, Statistics and Actuarial
Science of the University of Kent (Canterbury, United Kingdom) for the kind
hospitality during the realization of part of this work.
\end{acknowledgments}


\begin{thebibliography}{99}
\bibitem{n5} N. Manton and P. Sutcliffe, \textit{Topological Solitons}
(Cambridge University Press, Cambridge, England, 2004).

\bibitem{n4} E. Bogomol'nyi, Sov. J. Nucl. Phys. \textbf{24}, 449 (1976). M.
Prasad and C. Sommerfield, Phys. Rev. Lett. \textbf{35}, 760 (1975).

\bibitem{ano} H. J. de Vega and F. A. Schaposnik, Phys. Rev. D \textbf{14},
1100 (1976).

\bibitem{onshell} A. N. Atmaja, H. S. Ramadhan and E. da Hora, J. High
Energy Phys. \textbf{1602}, 117 (2016).

\bibitem{n1} H. B. Nielsen and P. Olesen, Nucl. Phys. B \textbf{61}, 45
(1973).

\bibitem{cshv} R. Jackiw and E. J. Weinberg, Phys. Rev. Lett. \textbf{64},
2234 (1990). R. Jackiw, K. Lee and E. J. Weinberg, Phys. Rev. D \textbf{42},
3488 (1990).

\bibitem{mcshv} C. Lee, K. Lee and H. Min, Phys. Lett. B \textbf{252}, 79
(1990).

\bibitem{loginov} A. Yu. Loginov, Phys. Rev. D \textbf{93}, 065009 (2016).

\bibitem{casana} R. Casana, M. L. Dias and E. da Hora, Phys. Lett. B \textbf{%
768}, 254 (2017).

\bibitem{cscp2} V. Almeida, R. Casana and E. da Hora, Phys. Rev. D \textbf{97%
}, 016013 (2018). R. Casana, M. L. Dias and E. da Hora, Phys. Rev. D \textbf{%
98}, 056011 (2018).

\bibitem{mcscp2} R. Casana, N. H. Gonzalez-Gutierrez and E. da Hora,
Europhys. Lett. \textbf{127}, 61001 (2019).

\bibitem{mcp2df} R. Casana, M. L. Dias and E. da Hora, Phys. Rev. D \textbf{%
96}, 076013 (2017).

\bibitem{mcp2is} J. Andrade, R. Casana, E. da Hora and C. dos Santos, Phys.
Rev. D \textbf{99}, 056014 (2019).

\bibitem{Shapoval:2010} T. Shapoval, V. Metlushko, M. Wolf, B. Holzapfel, V.
Neu and L. Schultz, Phys. Rev. B \textbf{81}, 092505 (2010).

\bibitem{Tung:2006} S. Tung, V. Schweikhard and E. A. Cornell, Phys. Rev.
Lett. \textbf{97}, 240402 (2006).

\bibitem{Anderson:1975zze} P.~W.~Anderson and N.~Itoh, Nature \textbf{256},
25 (1975).

\bibitem{Bulgac:2013nmn} A.~Bulgac, M.~M.~Forbes and R.~Sharma, Phys.\ Rev.\
Lett. \textbf{110}, 241102 (2013).

\bibitem{Wlazlowski:2016yoe} G.~Wlaz\l {}owski, K.~Sekizawa, P.~Magierski,
A.~Bulgac and M.~M.~Forbes, Phys.\ Rev.\ Lett.\ \textbf{117}, 232701 (2016).

\bibitem{Adam:2018pvd} C.~Adam and A.~Wereszczynski, Phys.\ Rev.\ D \textbf{%
98}, 116001 (2018).

\bibitem{Adam:2019yst} C.~Adam, J.~M.~Queiruga and A.~Wereszczynski, J. High
Energy Phys. \textbf{1907}, 164 (2019).

\bibitem{Goatham:2010dg} S.~W.~Goatham, L.~E.~Mannering, R.~Hann and
S.~Krusch, Acta Phys.\ Polon.\ B \textbf{42}, 2087 (2011).

\bibitem{Manton:1997tg} N.~S.~Manton, Annals Phys.\ \textbf{256}, 114 (1997).

\bibitem{Romao:2004df} N.~M.~Romao and J.~M.~Speight, Nonlinearity \textbf{17%
}, 1337 (2004).

\bibitem{Krusch:2005wr} S.~Krusch and P.~Sutcliffe, Nonlinearity \textbf{19}%
, 1515 (2006).

\bibitem{15} D. Tong and K. Wong, J. High Energy Phys. \textbf{1401}, 090
(2014).

\bibitem{16} X. Han and Y. Yang, Nucl. Phys. B \textbf{898}, 605 (2015). R.
Zhang and H. Li, Nonlin. Anal. \textbf{115}, 117 (2015).

\bibitem{18} X. Han and Y. Yang, J. High Energy Phys. \textbf{1602}, 046
(2016).

\bibitem{19} A. Cockburn, S. Krusch and A. A. Muhamed, J. Math. Phys.
\textbf{58}, 063509 (2017).

\bibitem{20} J. E. Ashcroft and S. Krusch, Phys. Rev. D \textbf{101}, 025004
(2020).

\bibitem{Aoki} K. Aoki, K. Sakakibara, I. Ichinose and T. Matsui, Phys. Rev.
B \textbf{80}, 144510 (2009).

\bibitem{CWu} C. Wu, J. Hu and S. Zhang, Phys. Rev. Lett. \textbf{91},
186402 (2003).

\bibitem{Stoof} H. T. C. Stoof, E. Vliegen and U. Al Khawaja, Phys. Rev.
Lett. \textbf{87}, 120407 (2001).

\bibitem{Chang} D. E. Chang, Phys. Rev. A \textbf{66}, 025601 (2002).

\bibitem{Kasamatsu} K. Kasamatsu, M. Tsubota and M. Ueda, Int. J. Mod. Phys.
B \textbf{19}, 1835 (2005).

\bibitem{fftt} Along the present manuscript, the metric signature is
considered as $\eta _{\mu \nu }=(+--)$. Moreover, Greek indices represent
space-time coordinates, while the Latin ones label spatial coordinates only.
Here, we also use the Natural Units System.

\bibitem{witten} E. Witten and D. Olive, Phys. Lett. B \textbf{78}, 97
(1978).

\bibitem{spector} Z. Hlousek and D. Spector, Nucl. Phys. B \textbf{370}, 143
(1992).
\end{thebibliography}
\end{document}